\documentclass[reprint,
floatfix,
superscriptaddress,
nofootinbib,
amsmath,amssymb,
aps,
prd,
longbibliography, widetext
]{revtex4-2}
\usepackage{slashed}
\usepackage{multirow}
\usepackage{booktabs}
\usepackage{tabularray}
\usepackage{graphicx}
\usepackage{siunitx}
\usepackage[normalem]{ulem}
\usepackage{float}
\usepackage[dvipsnames]{xcolor}
\usepackage{bbm}
\usepackage[colorlinks,allcolors=blue,linktocpage]{hyperref}
\usepackage{physics}
\usepackage[margin=false,inline=true,index=true,draft]{fixme}
\fxsetup{inlineface=\footnotesize}
\fxusetheme{colorsig}

\begin{document}
\title{Long-lived HNLs at lepton colliders as a probe of left-right-symmetric models}
\author{Kevin A.\ Urqu\'ia-Calder\'on}

\affiliation{Niels Bohr Institute, University of Copenhagen, Blegdamsvej 17, DK-2010, Copenhagen, Denmark}
\begin{abstract}
Left-right symmetric models (LRSM) were proposed to reconcile the apparent parity violation in weak interactions with our intrinsic notion of fundamental parity symmetry. It was quickly realized that LRSM offers a viable framework for explaining neutrino masses by incorporating right-handed neutrinos, \(N\), into its spectrum.
In this study, we investigate for the first time the potential of future lepton colliders (including the FCC-ee, CEPC, ILC, CLIC, and a muon collider) to detect signals stemming from \(N\)'s, focusing primarily on displaced vertex signals, and on prompt signals to a lesser degree.
Our results demonstrate that these signals can effectively probe the characteristic scales of LRSM, providing evidence that would be unobtainable through any other experimental means.
\end{abstract}
\maketitle

\section{Introduction}
Since the discovery of parity violation in weak interaction~\cite{Lee:1956qn, Wu:1957my}, scientists have been trying to reconcile this result with the intuitive expectation of a left-right symmetric world. Left-right symmetric models (LRSM) explored the idea that parity is restored at higher energy scales \cite{Pati:1974yy, Mohapatra:1974hk, Mohapatra:1974gc, Senjanovic:1975rk, Senjanovic:1978ev, Mohapatra:1979ia}. Such models, if a proper scalar sector is chosen, can also give rise to a seesaw mechanism which naturally gives neutrinos small masses \cite{Minkowski:1977sc, Mohapatra:1980yp}, in accordance with existing observations (see \cite{Cai:2017mow} for a recent review).
The particle content of the LRSM version we will consider contains the heavy cousins of intermediate vector bosons ($W_R$ and $Z^\prime$), right-handed neutrinos ($N$, heavy neutral leptons or HNLs) and an extended scalar sector, responsible for the left-right and electroweak symmetry breaking.
Owing to this, LRSM possess rich phenomenology and their observational signatures remain a valid goal for particle physics experiments.

The main LRSM search channel at the Large Hadron Collider (LHC) is the Keung-Senjanović process (\emph{KS process}) \cite{Keung:1983uu} -- the production of a heavy charged gauge boson which can result in a same-sign lepton pair plus jets, the collider equivalent of a neutrinoless double-beta decay. Both CMS \cite{CMS:2012zv, CMS:2016fxb, CMS:2017xcw, CMS:2018iye, CMS:2018agk, CMS:2021dzb} and ATLAS \cite{ATLAS:2015gtp, ATLAS:2018dcj, ATLAS:2019isd, ATLAS:2023cjo} have done extensive searches for this signal without avail.
The proposed high-intensity beam-dump experiments such as SHIP \cite{Alekhin:2015byh, Castillo-Felisola:2015bha, Mandal:2017tab}, MATHUSLA \cite{Dev:2017dui, Curtin:2018mvb} and FASER \cite{Kling:2018wct} may also provide complementary searches of LRSM.

Besides fixed target experiments, the particle physics community actively discusses the possibility of a lepton collider.
This includes both circular colliders such as the Future Circular Collider (FCC-ee) \cite{FCC:2018byv} and Circular Electron Positron Collider (CEPC) \cite{CEPCStudyGroup:2018rmc}, and linear colliders like the International Linear Collider (ILC) \cite{ILC:2007oiw}, and Compact Linear International Collider (CLIC) \cite{Aicheler:2012bya}. Additionally, the idea of a muon collider has recently gained significant traction \cite{Aime:2022flm, Black:2022cth, Accettura:2023ked}.
All these experiments will open new possibilities for searches of the LRSM signatures \cite{Rizzo:1981dla, Lusignoli:1989tr, Rizzo:1991ie, Maalampi:1991fx, Maalampi:1992np, Lepore:1994xw, Huitu:1997vh, Barry:2012ga}, including potential sensitivity for the extended scalar sector of the LRSM \cite{Dev:2017dui, Dev:2017ftk, Dev:2018sel, BhupalDev:2018tox, BhupalDev:2018vpr, Belfkir:2023lot}. However, there have been very few studies on the potential that future colliders offer on searches of HNLs in the LRSM.

In this work, we will provide a first study of the potential that proposals for lepton colliders for probing parameters in the LRSM. In particular, we will focus on the potential that displaced vertices from the production of two HNLs.
This collider signature appears if particles travel macroscopic distances (usually above a few \si{mm}) prior to decay. Such searches have the benefit of having little to no SM background, and, as we will show, have the benefit of probing scales of LRSM that are unreachable by current experiments.
The literature already offers us constraints from displaced HNLs at the LHC in the LRSM \cite{Nemevsek:2018bbt, Cottin:2018kmq}, but it does not on the potential of future lepton colliders for detecting long-lived HNLs (for the potential including HNLs at lepton collider proposals, see \cite{Blondel:2014bra, Antusch:2016ejd, Antusch:2016vyf, Barducci:2022hll, Mikulenko:2023ezx, Blondel:2022qqo} and references therein).

In the case of the LRSM, searches for displaced HNLs will be more sensitive \si{GeV}-scale HNLs, and to higher scale LRSM as opposed to what prompt signals can probe (the reason why will be detailed later). The region of the parameter space displaced vertices can search for is of particular interest when considering \si{keV}-scale HNLs as dark matter \cite{Bezrukov:2009th}. They require \si{GeV}-scale HNLs and very heavy gauge bosons to attain the observed dark matter density (which may be subject to strong constraints from large-scale structures \cite{Nemevsek:2022anh}) and which may ameliorate issues with the perturbativity of the theory \cite{Maiezza:2016bzp, Maiezza:2016ybz}. It can also probe regions of the parameter space where leptogenesis can be generated by LRSM \cite{Frere:2008ct} (or for a recent review on leptogenesis scenarios with HNLs see \cite{Hati:2018tge}).

In Sec.~\ref{sec:theory} we will review the necessary theoretical background of the LRSM. In Sec.~\ref{sec:constraints} we discuss the existing constraints and the sensitivity that future experiments will have on LRSM parameters. Sec~\ref{sec:colliders} reviews the current proposals for lepton colliders and their proposed parameters. Sec~\ref{sec:pheno} describes the production and decay channels for HNLs in lepton colliders and quickly discusses their signals, and in Sec.~\ref{sec:results} we present the expected sensitivity and discuss their implications. Finally, we conclude and summarize our results in Sec.~\ref{sec:conclusions}. The Appendixes provide additional calculations considering the effects that the $W-W_R$ mixing and that leptonic mixing may have on our results, and briefly touch on the sensitivity of prompt searches in the most optimistic of cases.

\section{The minimal Left-Right Symmetric Model} \label{sec:theory}
Left-Right symmetric models (LRSM) are based on the following gauge group:
\begin{equation}
    \mathrm{SU}(2)_R \times \mathrm{SU}(2)_L \times \mathrm{U}(1)_{B-L}\,,
\end{equation}
which is coupled with a discrete generalized charge conjugation $\mathcal{C}$ or a parity conjugation $\mathcal{P}$ (see \cite{Maiezza:2010ic} for a precise difference between both cases).  Both restrict the number of free parameters, like the gauge couplings of the left- and right-handed $\mathrm{SU}(2)$ that sets $g_L = g_R$ at tree level, making the model more predictive and renders the particle content left-right symmetric.

The fermion sector consists of both left- and right-handed doublets:
\begin{equation}
    \label{eq:fermionic_doublets}
    Q_{L,R} = \begin{pmatrix}
    u \\ d
    \end{pmatrix}_{L,R}\,, \hspace{1cm}
    L_{L,R} = \begin{pmatrix}
    \nu \\ \ell
    \end{pmatrix}_{L,R}\,.
\end{equation}
Moreover, the additional $\mathrm{SU}(2)_R$ gauge group will give us more gauge bosons--a charged gauge boson and a neutral gauge boson. The charged gauge bosons with a defined mass will be denoted as $W_R^\pm$ (as opposed to the SM $W$ boson), and the neutral one will be as $Z^\prime$ (as opposed to the $Z$).

We work in the case where we have two triplets $\Delta_{L,R}$ and a bidoublet $\Phi$,

\begin{equation}
    \label{eq:scalars}
    \Delta_{L,R} = \begin{pmatrix}
        \dfrac{\delta_{L,R}^+}{\sqrt{2}} & \delta_{L,R}^{++} \\
        \delta_{L,R}^0 & -\dfrac{\delta_{L,R}^+}{\sqrt{2}}
    \end{pmatrix}\,, \hspace{0.5cm}
    \Phi = \begin{pmatrix}
        \phi_1^0 & \phi_2^+ \\
        \phi_1^- & -\phi_2^{0*}
    \end{pmatrix}\,.
\end{equation}
These scalars will have the following vacuum expectation values (vev's):
\begin{equation}
    \label{eq:vevs}
    \left\langle\delta_{L,R}^0 \right\rangle = \frac{v_{L,R}}{\sqrt{2}}\,,\hspace{0.5cm} \left\langle\Phi\right\rangle = \frac{v}{\sqrt{2}}\,\mathrm{diag}(\cos\beta, -\sin\beta e^{-i\alpha})\,,
\end{equation}
where $v_R$ will be the scale at which the LRSM gauge group is broken into the SM gauge group, and $v$ is the usual SM vev. $\beta$ and $\alpha$ are arbitrary parameters related to the mixing of the vevs of the bidoublets. The theory has no way of predicting the value of these parameters but in the case of $\mathcal{P}$, we can derive indirect constraints from the Hermiticity of quark mass matrices \cite{Maiezza:2010ic, Senjanovic:2014pva, Senjanovic:2015yea}, and from the QCD parameter $\theta$ \cite{Maiezza:2014ala, Dekens:2021bro}. The value of $v_L$ is expected to be much smaller than both $v$ and $v_R$ because it directly contributes to neutrino masses.

The breaking of the $\mathrm{SU}(2)_R$ group will give masses to the right-handed gauge bosons of the LRSM, and the subsequent breaking of the SM group will give masses to the rest of the gauge bosons. The entire mass spectrum of the gauge bosons is
\begin{equation}
    \begin{aligned}
        \label{eq:mass_spectrum_gauge}
        M_{W}^2 &\simeq \frac{1}{4}g^2 v^2\,,        & M_{W_R}^2 &\simeq \frac{1}{2}\,g^2 v_R^2\,, \\
        M_{Z}^2 &\simeq \frac{1}{4 c_w^2} g^2 v^2\,, & M_{Z^\prime}^2 &\simeq g^2 v_R^2 \frac{c_w^2}{c_{2w}}\,, & m_{\gamma}^2 &= 0\,,
    \end{aligned}
\end{equation}
where $c_w \equiv \cos\theta_w$ is the cosine of the Weinberg angle. Just as the SM predicted that the mass of the $Z$ boson is heavier than the mass of the $W$, the LRSM predicts $M_{Z_R} \simeq 1.69 M_{W_R}$.

Furthermore, the model will also give rise to active neutrino mass terms which will mix with right-handed neutrinos, $N_R$, (which we will call \textit{heavy neutral leptons} or HNLs) in a sort of seesaw mechanism \cite{Mohapatra:1980yp}. The mass terms are:
\begin{equation}
    \label{eq:see_saw_lagrangian}
    \mathcal{L}_{\nu N} = 
    -\frac{1}{2} \begin{pmatrix} \bar{\nu}_L &  \bar{N}_R^C \end{pmatrix} 
    \begin{pmatrix}
        M_L^\dagger & M_D \\[2ex] M_D^T & M_R 
    \end{pmatrix}
    \begin{pmatrix} \nu_L^C \\[2ex]  N_R \end{pmatrix} + \mathrm{H.c.}\,,
\end{equation}
here $M_L$ is proportional to $v_L$, $M_D$ to $v$ and $M_R$ to $v_R$, all up to a Yukawa coupling. A unitary transformation will diagonalize the mass matrix. This unitarity matrix will depend on the mixing angle matrix $\Theta$. The mass spectrum of this matrix will give us Majorana mass terms for both active neutrinos and HNLs, the former will be naturally small as a feature of the difference of scales.

The mixing angle $\Theta$ will induce new interactions between HNLs and the usual SM particles (as well as interactions between active neutrinos and the right-handed sector), but in this model, $\Theta$ will be very small ($\Theta \propto \sqrt{m_\nu/m_N}$, where $m_N$ is the mass of the HNLs) \cite{Nemevsek:2012iq, Senjanovic:2016vxw} and therefore, not considered in any of our analyses. 

The model has no way of predicting the mass spectrum or the masses of HNLs. For simplicity, we will work in the limit where two HNLs are decoupled and only one HNL will be accessible at collider energies.

\begin{figure*}[t]
    \centering
    \includegraphics[width = \linewidth]{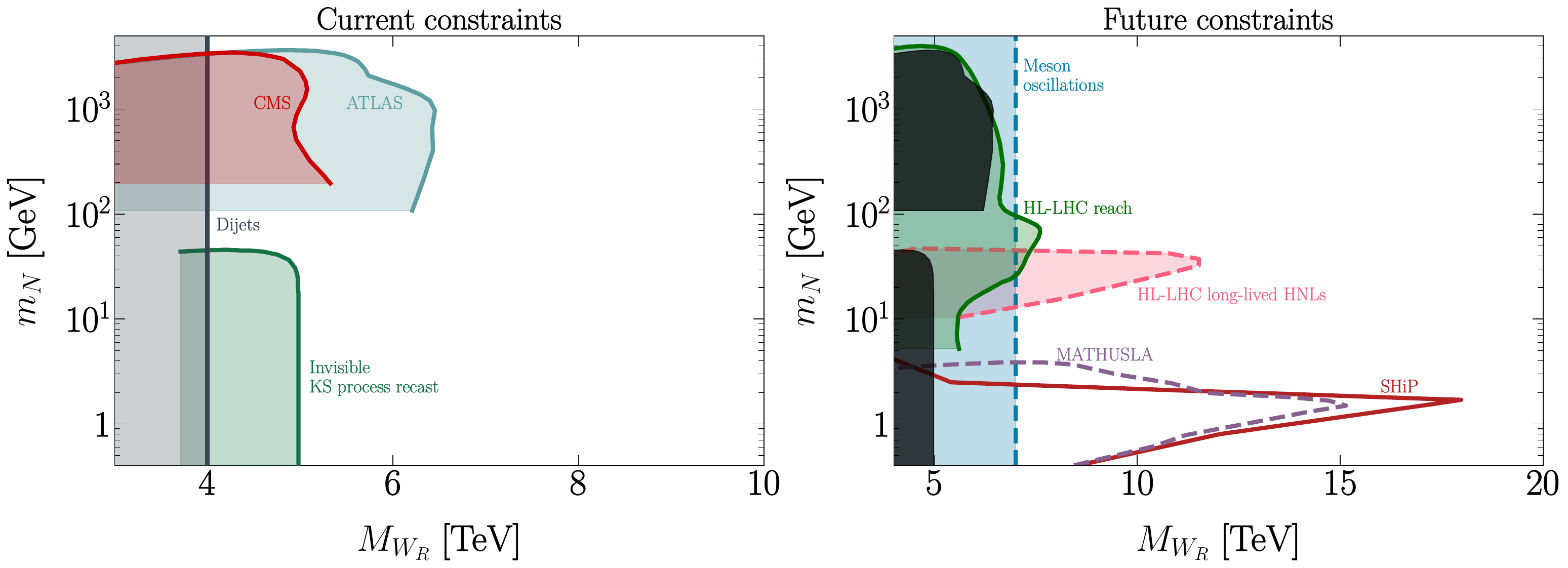}
    \caption{Constraint summary on a $m_N$ and $M_{W_R}$ parameter space plot. The left (right) plot shows the constraints from current (future) experiments. We show the current constraints from ATLAS \cite{ATLAS:2023cjo}, CMS \cite{CMS:2021dzb}, dijets signals \cite{ATLAS:2019fgd, CMS:2019gwf} which can interpreted a $W_R$ decaying to a jet pair, and signals of high-energy leptons with missing energy \cite{CMS:2018hff, ATLAS:2019lsy} which can be interpreted as a KS-process with an HNL so light it decays outside of the detector \cite{Nemevsek:2018bbt}. We also show the prospective bounds from the HL-LHC \cite{Nemevsek:2018bbt}, a KS process with a displaced vertex \cite{Cottin:2019drg}, meson oscillations \cite{Bertolini:2014sua}, SHIP \cite{Alekhin:2015byh, Castillo-Felisola:2015bha, Mandal:2017tab} and MATHUSLA \cite{Dev:2017dui, Curtin:2018mvb}.}
    \label{fig:constraints}
\end{figure*}

\subsection{Interaction terms from the Lagrangian}
We will now present the interaction terms from the Lagrangian that will be relevant for the production of HNLs at lepton colliders.\footnote{For a complete overview of all the interactions in the Lagrangian, the reader can check \cite{Roitgrund:2014zka}.}
\begin{align}
    \label{eq:W_interactions_N}
    \mathcal{L}_{W\ell N} &\simeq \frac{g}{\sqrt{2}} U_{\alpha i}\, \bar{\ell}_\alpha \slashed{W}_R^-\,P_R N_i - \frac{g}{\sqrt{2}} U_{\alpha i}\,\xi_W \bar{\ell}_\alpha \slashed{W}^-\,P_R N_i\,,
    \\
    \label{eq:Z_interactions_N}
    \mathcal{L}_{ZN} &\simeq  \frac{g\,c_w}{2\sqrt{c_{2w}}} \bar{N}_R \slashed{Z}^\prime N_R - \frac{g\,c_w}{2\sqrt{c_{2w}}} \zeta_Z \bar{N}_R \slashed{Z} N_R\,,
\end{align}
where $\xi_W$ and $\zeta_Z$ are:
\begin{align}
    \label{eq:mixing_W}
    \xi_W   &= \frac{M_{W}^2}{M_{W_R}^2} \sin(2\beta) e^{i \alpha}\,,\\
    \label{eq:mixing_Z}
    \zeta_Z &= -\frac{c_{2w}^{3/2}}{2 c_w^4} \frac{M_{W}^2}{M_{W_R}^2}\,,
\end{align}
and where $U_{\alpha i}$ are the matrix elements of the unitary mixing matrix of HNLs, analogous to the PMNS matrix.
For our analysis, we will consider two different shapes of $U$:

\begin{align}
\label{eq:right-handed-PMNS-parametrizations}
    U^{e} &= \begin{pmatrix}
        1 & 0 & 0 \\ 0 & 1 & 0 \\ 0 & 0 & 1 
    \end{pmatrix}\,, & 
    U^{\mu} &= \begin{pmatrix}
        0 & 1 & 0 \\ 1 & 0 & 0 \\ 0 & 0 & 1 
    \end{pmatrix}\,. 
\end{align}
As mentioned before, we will consider only one HNL to be reachable within collider energies, whereas the other two are too massive to be produced. Here, we will consider $N_1$ to be that accessible HNL, which means that in the case of $U^e$, then our HNL will only interact with electrons; in the second case, $U^\mu$ it will only interact with muons.

Our final results will depend on the shape of $U$. 
In Appendix~\ref{app:different_mixing}, we show how much our results would change for an arbitrary shape of $U$ and then consider some specific cases.

In this model, HNLs will primarily decay semileptonically. For the sake of completeness, we will also include the relevant interaction term that will mediate such decays
\begin{equation}
    \label{eq:W_interactions_quarks}
    \mathcal{L}_{W_RUD} = \frac{g}{\sqrt{2}} V_{UD} \bar{U}_R \slashed{W}_R^- D_R + \mathrm{H.c.}\,,
\end{equation}
where $V$ is the usual CKM matrix, and $U$, $D$ represent any up and down quark type respectively. In general, the \textit{right-handed} and the \textit{left-handed} CKM matrices are not necessarily equal; there will be a discrepancy in \textit{CP} phases between both (see \cite{Maiezza:2010ic}), which should not affect our results.

\section{Constraint summary} \label{sec:constraints}
In this section, we will present a summary of experimental and theoretical constraints on LRSM parameters. We summarize some of these in Fig.~\ref{fig:constraints}. We will later compare our final results with the current and prospective bounds found in the literature.

\paragraph{Constraints from hadron colliders.} The signals in which the LRSM can manifest itself in hadron colliders have been well studied in the past \cite{Keung:1983uu, Ferrari:2000sp, Maiezza:2010ic, Das:2012ii, Nemevsek:2018bbt}. The main channel searched for is the Keung-Senjanović process (KS process) ($pp \to W_R \to [N]\ell \to [jj\ell] \ell$). This process can result in the production of two final same-sign leptons due to the Majorana nature of HNLs, which subsequently makes this channel have no irreducible SM background. This channel depends on the mass of the HNL and $W_R$. Both ATLAS \cite{ATLAS:2015gtp, ATLAS:2018dcj, ATLAS:2019isd, ATLAS:2023cjo} and CMS \cite{CMS:2012zv, CMS:2016fxb, CMS:2017xcw, CMS:2018iye, CMS:2018agk, CMS:2021dzb} have performed numerous searches for this channel. We can find an analysis of the sensitivity of the high luminosity LHC (HL-LHC) in \cite{Nemevsek:2018bbt}.

Both ATLAS and CMS have also performed searches for \textit{heavier $W$ bosons} by searching for pairs of highly energetic jets \cite{ATLAS:2019fgd, CMS:2019gwf} and for a lone very high energy charged lepton and missing transverse energy \cite{ATLAS:2019lsy, CMS:2018hff}. The former can be interpreted as a $W_R$ decaying to a pair of quarks and would give us constraints solely of the mass of $W_R$, and the latter to it decaying into a charged lepton and a very long-lived HNL \cite{Nemevsek:2018bbt}.

Neither ATLAS nor CMS has searched for a KS process with a long-lived HNL; they have only searched for prompt HNLs. The sensitivity for these channels was presented in the past \cite{Helo:2013esa, Nemevsek:2018bbt, Cottin:2019drg} and is particularly sensitive to lighter HNLs and heavier $W_R$ when compared to the prompt-HNL searches.

We should emphasize that CMS, ATLAS, and LHCb have all performed searches for long-lived HNLs, but not in the LRSM, but in the minimal type-I seesaw \cite{ATLAS:2022atq, ATLAS:2019kpx, CMS:2022fut}. A reinterpretation of these analyses may not be entirely feasible because they were done for leptonic decays of HNLs and, as we will see later on, HNLs in this model decay primarily semileptonically.

\paragraph{Constraints from indirect physics.} The new particles from the LRSM affect a plurality of different low-energy physics phenomena, including meson mixing \cite{Ecker:1985vv, Bertolini:2014sua}, the neutron electron dipole (nEDM) in the case of $\mathcal{P}$ \cite{Georgi:1978xz, Beg:1978mt, Ecker:1985vv, Babu:1988mw, Babu:1989rb, Barr:1991qx, Maiezza:2014ala, Bertolini:2019out, Dekens:2021bro}, $0\nu\beta\beta$ decay,  as well as electroweak precision observables \cite{Hsieh:2010zr, Bernard:2020cyi}. 

Meson mixing provides a lower bound of $M_{W_R} \gtrsim \SI{3}{TeV}$, and the next runs of Belle II and LHCb will provide bounds up to $M_{W_R} \gtrsim \SI{7}{TeV}$ \cite{Bertolini:2014sua}. These future bounds will be stronger than the constraints the LHC can give on a dijet signal.

Constraints from nEDM provide limits as small as $M_{W_R} \gtrsim \SI{17}{TeV}$ which could be ameliorated by including a Peccei-Quinn mechanism, down to $M_{W_R} \gtrsim \SI{5.5}{TeV}$ \cite{Dekens:2021bro}.

\paragraph{Constraints from high-intensity experiments.} Future high-intensity experiments like SHIP \cite{Alekhin:2015byh}, and MATHUSLA \cite{Curtin:2018mvb} will provide complementary constraints to the ones from hadron colliders. High-intensity experiments aim to search for displaced vertices from long-lived particles and will be particularly sensitive to HNLs with lower masses and $W_R$ with higher masses than the ones from hadron colliders. Both SHIP \cite{Alekhin:2015byh, Castillo-Felisola:2015bha, Mandal:2017tab} and MATHUSLA \cite{Dev:2017dui, Curtin:2018mvb} which have sensitivities up to $M_{W_R} \simeq \SI{18}{TeV}$ and $m_N \simeq \SI{5}{GeV}$.

\paragraph{Theoretical constraints.}
A study on vacuum stability dictated that HNLs cannot be heavier than the mass of $W_R$ \cite{Mohapatra:1986pj}, another study that considered the perturbativity of the theory found less restrictive constraints allowing HNLs to be heavier than $W_R$ \cite{Maiezza:2016bzp}. As we will see later, our results fall in a region of parameter space allowed by both studies. For studies regarding the perturbativity and stability of the scalar sector, we direct the interested reader to the pertinent literature \cite{Olness:1985bg, Chakrabortty:2016wkl, Maiezza:2016bzp, Maiezza:2016ybz, Chauhan:2018uuy}.

We summarize most of the experimental constraints in  Fig.~\ref{fig:constraints}.

\section{Proposals for new particle colliders} \label{sec:colliders}
The HL-LHC is scheduled to cease operations by December 2041. The particle physics community has actively discussed the possibility of building a new collider as the next step in collider physics. The current consensus favors a new lepton collider. Lepton colliders have the advantage of having lower uncertainties compared to hadron colliders: they allow us to do better precision measurements. However, it is harder for them to explore higher energies. Currently, there are two types of proposals: circular and linear colliders: 

\begin{itemize}
    \item \textbf{FCC}: the Future Circular Collider \cite{FCC:2018byv, FCC:2018evy} is a proposed ~\SI{90}{km} circular collider. It is currently the most popular option for physicists in Europe. It is meant to be built at CERN, between Switzerland and France. The first stage, FCC-ee, where electron-positron pairs will collide, will produce an impressive $10^{12}$ $Z$ bosons, $10^8$ $W$'s, $10^6$ Higgs bosons and $10^6$ top quarks. After this first stage, it will be upgraded to hadron collisions and later to electron-hadron collisions. The high intensity of the FCC-ee makes it such an attractive machine for new physics searches, both direct and indirect.
    \item \textbf{CEPC}: the Circular Electron-Positron Collider \cite{CEPCStudyGroup:2018rmc, CEPCStudyGroup:2018ghi} is a ~\si{100}{km} electron-positron collider in China meant to begin operations during the 2030s. It will run at the same energies as the FCC-ee, albeit with smaller luminosities. A subsequent upgrade to a proton-proton collider is also expected. 
    \item \textbf{ILC}: the International Linear Collider \cite{ILC:2007oiw} is a proposed electron-positron linear collider which will stretch approximately \SI{20}{km} in length. If built, the most likely host would be either in Japan or CERN. Its proposed energies are higher than that of the FCC-ee and CEPC, potentially up to $\sqrt{s} \simeq \SI{500}{GeV}$.
    \item \textbf{CLIC}: the Compact Linear Collider \cite{Aicheler:2012bya, CLICdp:2018cto} is a proposed electron-positron linear with a length stretching between \si{11}--\SI{50}{km}. It is another alternative for CERN in case the FCC is not built. Its proposed energy range surpasses that of the ILC. It hopes to reach up to \SI{3}{TeV}. 
    \item \textbf{Muon collider}: the proposal for a muon-collider has received plenty of attention recently \cite{Aime:2022flm, Black:2022cth, Accettura:2023ked}. Muon colliders possess the potential to explore higher center-of-mass energies compared to electron colliders, primarily due to muons experiencing fewer energy losses from synchrotron radiation. Furthermore, as muons are elementary particles, the entire beam energy can be utilized for collisions, which sets them apart from proton colliders. Suggestions for a muon-muon collider claim it may reach up to $\sqrt{s} \simeq \SI{10}{TeV}$. Most of the attention is receiving is from physicists in the U.S., and it seems like a likely place for it to be built, but it may remain a possibility for CERN in case the FCC is not built. Despite how attractive a muon collider may seem, several studies and technological advances are needed before it can even come close to becoming a reality.
\end{itemize}

The proposed center of mass energies and luminosities are all summarized in Table~\ref{tab:energies_luminosities}.

\begin{table}[ht]
    \centering
    \begin{tabular}{|c|c|c|}
        \hline
        Experiment & Energy (\si{GeV}) & Luminosity (\si{ab^{-1}}) \\ \hline
        FCC-ee &  \numlist{90;161;240;350} & \numlist{150;10;5;1.7} \\
        CEPC & \numlist{90;161;240;360} & \numlist{100;6;20;1} \\
        ILC &  \numlist{250;350;500} & \numlist{2;0.2;4} \\
        CLIC & \numlist{380;1500;3000} & \numlist{1;2.5;5} \\
        Muon collider & \numlist{3000;10000} & \numlist{0.4;4} (per year) \\
        \hline
    \end{tabular}
    \caption{Proposed luminosities and energies at which the FCC \cite{FCC:2018evy}, CEPC \cite{CEPCPhysicsStudyGroup:2022uwl}, ILC \cite{Barklow:2015tja, Bambade:2019fyw} and CLIC \cite{Aicheler:2018arh} may run at. A single proposal for a muon collider is not yet properly given, the points use will be used as benchmarks for our plots and were taken from \cite{Black:2022cth}.
    \label{tab:energies_luminosities}}
\end{table}

\section{HNL phenomenology in lepton colliders} \label{sec:pheno}
\begin{figure}[t]
    \centering
    \includegraphics{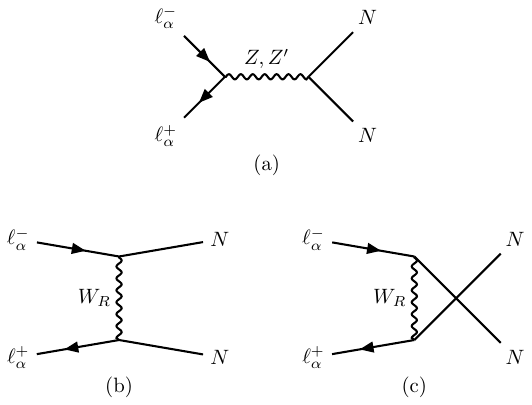}
    \caption{Dominating production in lepton colliders for HNL pairs}
    \label{fig:production_channels}
\end{figure}
As we stated before in the Introduction, ideas for searching for LRSM at lepton collider can already be found in the literature (for a summary see \cite{Huitu:1997vh}), including the pair production of charged gauge bosons (either $W_R, W_R$ or $W, W_R$) \cite{Maalampi:1992np, Maalampi:1992nq, Barry:2012ga}, the production and effects of double-charged scalars \cite{Rizzo:1981dla, Lusignoli:1989tr, Rizzo:1991ie, Lepore:1994xw, Dev:2018sel, BhupalDev:2018tox, BhupalDev:2018vpr}, neutral scalars \cite{Dev:2017ftk, BhupalDev:2018vpr}, as well as the idea we re proposing here: the production of two HNLs \cite{Ma:1989jpa, Kogo:1991ec, Maalampi:1991fx, Buchmuller:1991tu, Buchmuller:1992wm, Gluza:1993gf, Gluza:1994ac, Hofer:1996cs, Biswal:2017nfl}.

\subsection{Production}
There are four different production channels for a pair of HNLs, $W_R$ mediated $t$ and $u$ channels, and $s$ mediated $Z$ and $Z^\prime$ channels shown in Fig.~\ref{fig:production_channels}. The interactions between the heavy gauge bosons and HNLs are shown in Eqs.~\eqref{eq:W_interactions_N} and \eqref{eq:Z_interactions_N}. We can see that the $Z$ mediated channel is only possible due to the $\zeta_Z$ mixing. Both $ s$ channels will be dominant at their respective peaks, but the $W_R$ mediated channels will be dominant for any other center-of-mass energy. We computed the tree level cross section in \textit{Mathematica} with the help of \texttt{FeynCalc} \cite{Shtabovenko:2020gxv} and \texttt{FeynArts} \cite{Hahn:2000kx} with the implementation of LRSM from \cite{Roitgrund:2014zka}.

\begin{figure}[t]
    \centering
    \includegraphics[width = \linewidth]{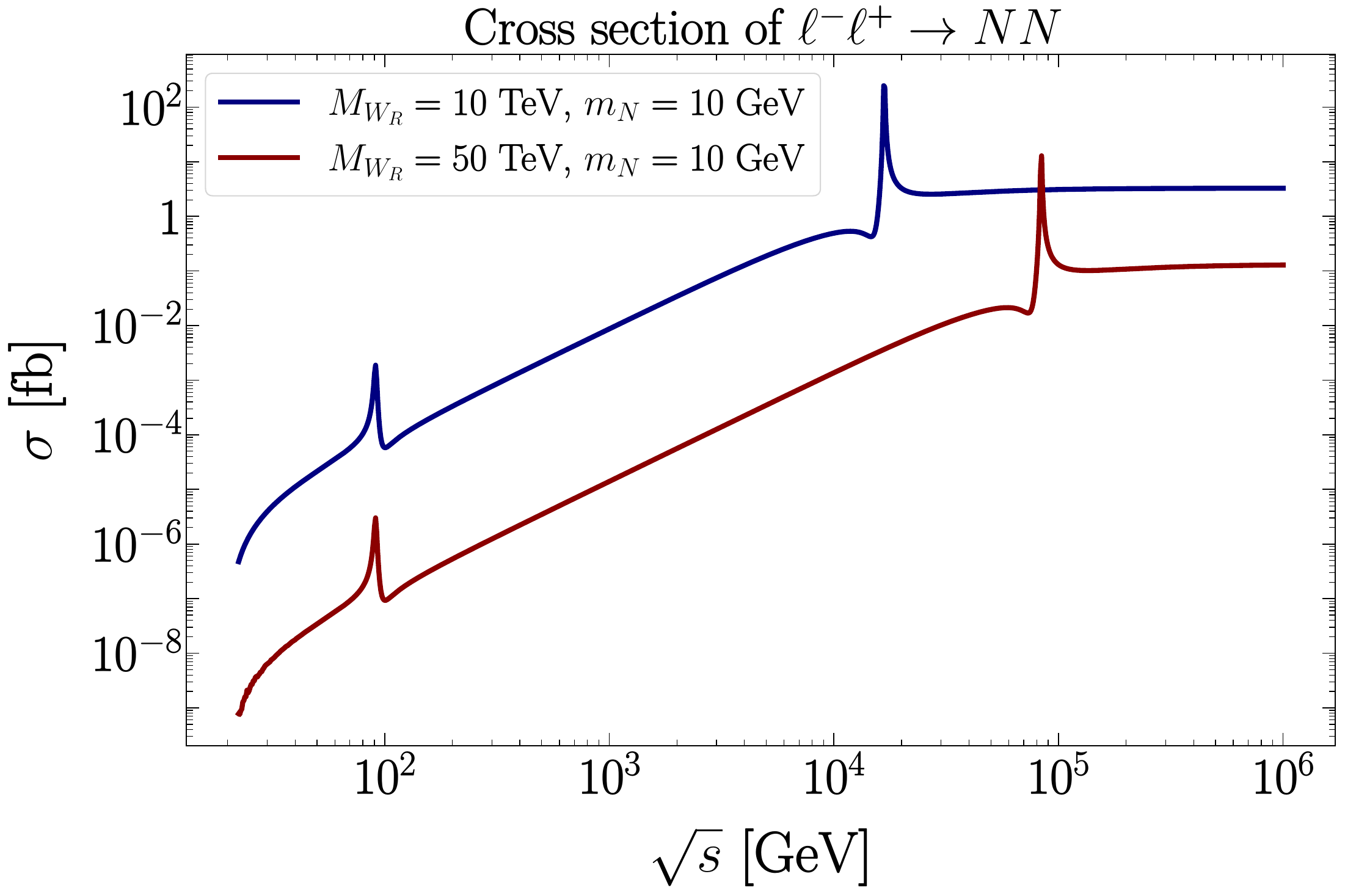}
    \caption{The total cross section for the production of a pair of HNLs in lepton-colliders, for $m_N = \SI{10}{GeV}$ and for values of $M_{W_R}$ shown in the plot. The two peaks correspond to the $Z$ and $Z^\prime$ peaks, the mass of $Z^\prime$ is $M_{Z^\prime} \simeq 1.69 M_{W_R}$ as in Eq.~\ref{eq:mass_spectrum_gauge}.}
    \label{fig:cross_section}
\end{figure}

At tree level, the cross section behaves as
\begin{equation}
    \label{eq:cross_section_behaviour}
    \sigma(\ell_\alpha^- \ell_\alpha^+\to NN) \propto \left\{
    \begin{aligned}
        &\frac{\xi_Z^2}{\Gamma_Z^2} & & \text{for } \sqrt{s} \simeq M_Z\,, \\[1.5ex]
        &\frac{1}{\Gamma_{Z^{\prime}}^2} & & \text{for } \sqrt{s} \simeq M_{Z^\prime}\,, \\[1.5ex]
        &\abs{U_{\alpha 1}}^4\,\frac{s}{M_{W_R}^4} & & 
        \begin{aligned}
            \text{for } &\sqrt{s} \neq M_Z, M_{Z^\prime} \\ 
            &\sqrt{s} \ll M_{W_R}\,, 
        \end{aligned} \\[1.5ex]
        &\frac{\abs{U_{\alpha 1}}^4}{M_{W_R}^2} & & \text{for } \sqrt{s} \gg M_{W_R}\,.
    \end{aligned}
    \right.
\end{equation}

The tree-level analytic expression of the cross section, as a function of $\sqrt{s}$, is plotted in Fig.~\ref{fig:cross_section} for different values of $M_{W_R}$. 

The first two lines are due to the dominance of the $s$ channels are their respective poles of Eq.~\eqref{eq:cross_section_behaviour}. The third and fourth lines are due to the $t$ and $u$ channel production. These contributions are all suppressed to a power of $M_{W_R}$, in the case of the $Z$ pole due to the $\zeta_Z$ mixing [shown in Eq.~\eqref{eq:mixing_Z}], and in the case of the $Z^\prime$ pole because the decay width of $Z^\prime$ depends on $M_{Z^\prime}$ (its value can be found in the Appendix~\ref{app:decay_width_Z}).

The third line in Eq.~\eqref{eq:cross_section_behaviour} indicates that the cross section increases quadratically with $\sqrt{s}$ when $\sqrt{s}$ is much smaller than the mass of $W_R$. This growth in cross section will dramatically increase the sensitivity for higher and higher energies. Indeed, the sensitivity of a $\sqrt{s} = \SI{10}{TeV}$ collider will be 100 times more powerful than the cross section of $\sqrt{s} = \SI{1}{TeV}$ collider with the same luminosity.

Of course, the cross section cannot grow arbitrarily with $\sqrt{s}$ as one would expect Yang-Mills theories with spontaneous symmetry breaking \cite{LlewellynSmith:1973yud, Cornwall:1973tb, Cornwall:1974km}. In the limit where $\sqrt{s} \gg M_{W_R}$ in Eq.~\eqref{eq:cross_section_behaviour}, the cross section plateaus. Figure~\ref{fig:cross_section} illustrates the overall behavior of the cross section.

We remind the reader that we will be working in the case where only one of the HNLs, $N_1$, is accessible at collider energies, and the other two are too massive. We will also only be considering the parametrizations of $U$ in Eq.~\eqref{eq:right-handed-PMNS-parametrizations}, where either $\abs{U_{e1}}^2 = 1$, or $\abs{U_{\mu 1}}^2 = 1$, the former maximizes the production cross section at linear colliders, whereas the latter at a muon collider.

If we choose a different parametrization for the $U$ matrix, then the production cross section will be affected, in particular for colliders that operate at energies where the main production channel will be through the intermediate $W_R$ in the $t$ or $u$ channel, where it will be the case for linear and muon colliders.

As it was noted beforehand in the literature \cite{Biswal:2017nfl}, the cross section can also enjoy an enhancement if we polarize the beams adequately. Both of the linear colliders, ILC and CLIC, will have their electron-positron beams polarized \cite{Moortgat-Pick:2005jsx, ILC:2007oiw, Aicheler:2012bya, Bambade:2019fyw}. If electrons are mostly right-handed and positrons left-handed then there will be an enhancement in cross section, but if it were the opposite we would have a diminution. This is because $W_R$ mostly interacts with right-handed electrons and left-handed positrons.\footnote{Interaction between $W_R$ and left-handed electrons and right-handed positrons are possible, but are suppressed due to the $\xi_W$ mixing in Eq.~\eqref{eq:mixing_W} or due to the very small $\Theta$ parameter.}

There are also other possibilities for the production of HNLs beyond what we have already presented. One is the production of only HNL instead of two ($\ell^- \ell^+ \to N \nu$), either through the $\nu-N$ mixing or through the $W-W_L$ one. As we have mentioned before, the former will be suppressed, but the later will depend on $\xi_W$ in Eq.\eqref{eq:mixing_W}, which depends on the arbitrary parameter $\beta$. However, the cross section of this channel is not competitive, this is elaborated upon in Appendix~\ref{app:production_beta}.

Another possibility would be to produce them through an intermediate scalar neutral scalar, either through direct production of the scalar or through mixing with the Higgs boson, as it was explored in \cite{Maiezza:2015lza}. Searches for HNLs from these production channels may lead to interesting results but will not be discussed any further in this paper.

\subsection{Decay}
\begin{figure}[t]
    \centering
    \includegraphics[width = \linewidth]{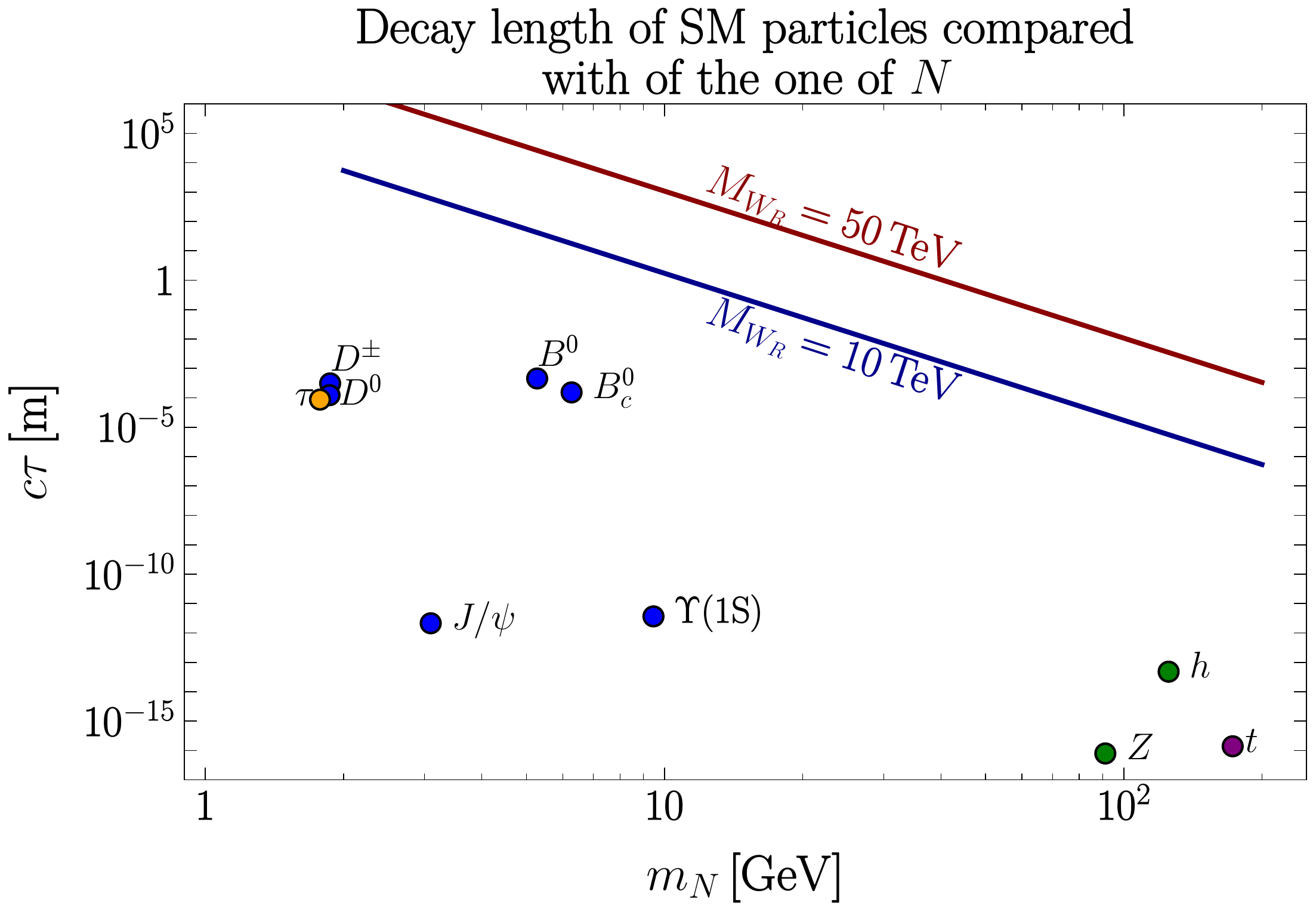}
    \caption{Decay length of $N$ in its own frame of reference for different values of $M_{W_R}$, compared with the decay length of other SM particles in their own frame of reference.}
    \label{fig:decay_length}
\end{figure}

HNLs will primarily decay through $W_R$ in this model. For the sake of simplicity, we will consider the channel $N \to \ell W_R^* \to \ell U D$, where $U, D$ can be any up and down-type quark or antiquark respectively. We can directly see these decays being possible from Eqs.~\eqref{eq:W_interactions_N} and \eqref{eq:W_interactions_quarks}. In the limit where quark and charged lepton masses are negligible compared to HNL masses and negligible transferred momentum, we have:
\begin{equation}
    \label{eq:decay_width_approximation} 
    \begin{aligned}[t]
        \Gamma_{\mathrm{tot.}} &\simeq \frac{G_F^2 m_N^5}{32 \pi^3} \frac{M_{W}^4}{M_{W_R}^4} \sum_{u,c;d,s} \abs{V_{UD}}^2\, \sum_{\alpha = e,\mu,\tau} \abs{U_{\alpha 1}}^2 \\
        &\simeq \frac{G_F^2 m_N^5}{16 \pi^3} \frac{M_{W}^4}{M_{W_R}^4}\,.    
    \end{aligned}
\end{equation}
The limit considered will be good enough for the region of parameter space in which we are working. Moreover, it justifies the usage of the NWA (narrow-width approximation), since
\begin{equation} \label{eq:decay_width_mass_ratio}
    \frac{\Gamma_\mathrm{tot.}}{m_N} \simeq \num{1e-5}\,\left(\frac{m_N}{M_{W_R}}\right)^4 \ll 1\,.
\end{equation}

For our specific choices in the parametrizations of $U$, then $N$ will only decay to a specific lepton, either an electron (or positron) for the case of $U^e$ and a muon (or antimuon) for the case of $U^\mu$. Different parametrization can allow for decays into possibly all charged leptons.

The lighter $m_N$ is and the heavier $M_{W_R}$ is, the lifetime of $N$ increases, which would lead to them being long-lived, as shown in Fig.~\ref{fig:decay_length}. 

In the case we have a non-negligible $W-W_R$ mixing, then the total decay width in Eq.~\eqref{eq:decay_width_approximation} changes due to having the possibility of $N$ decaying through an intermediate $W$ boson. This would change the decay length and lifetime of $N$, particularly for an $N$ heavier than a $W$ boson. This is elaborated upon in Appendix~\ref{app:decay_beta}.

\subsection{Signals}
\begin{figure}
    \centering
    \includegraphics[width = 0.9\linewidth]{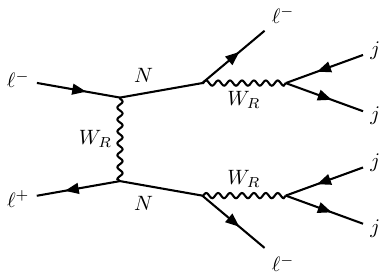}
    \caption{LNV process $\ell^- \ell^+ \to 2\ell^- 4j$.}
    \label{fig:LNV_diagram}
\end{figure}

\begin{figure*}[t]
    \begin{minipage}{0.45\textwidth}
        \includegraphics[width = \linewidth]{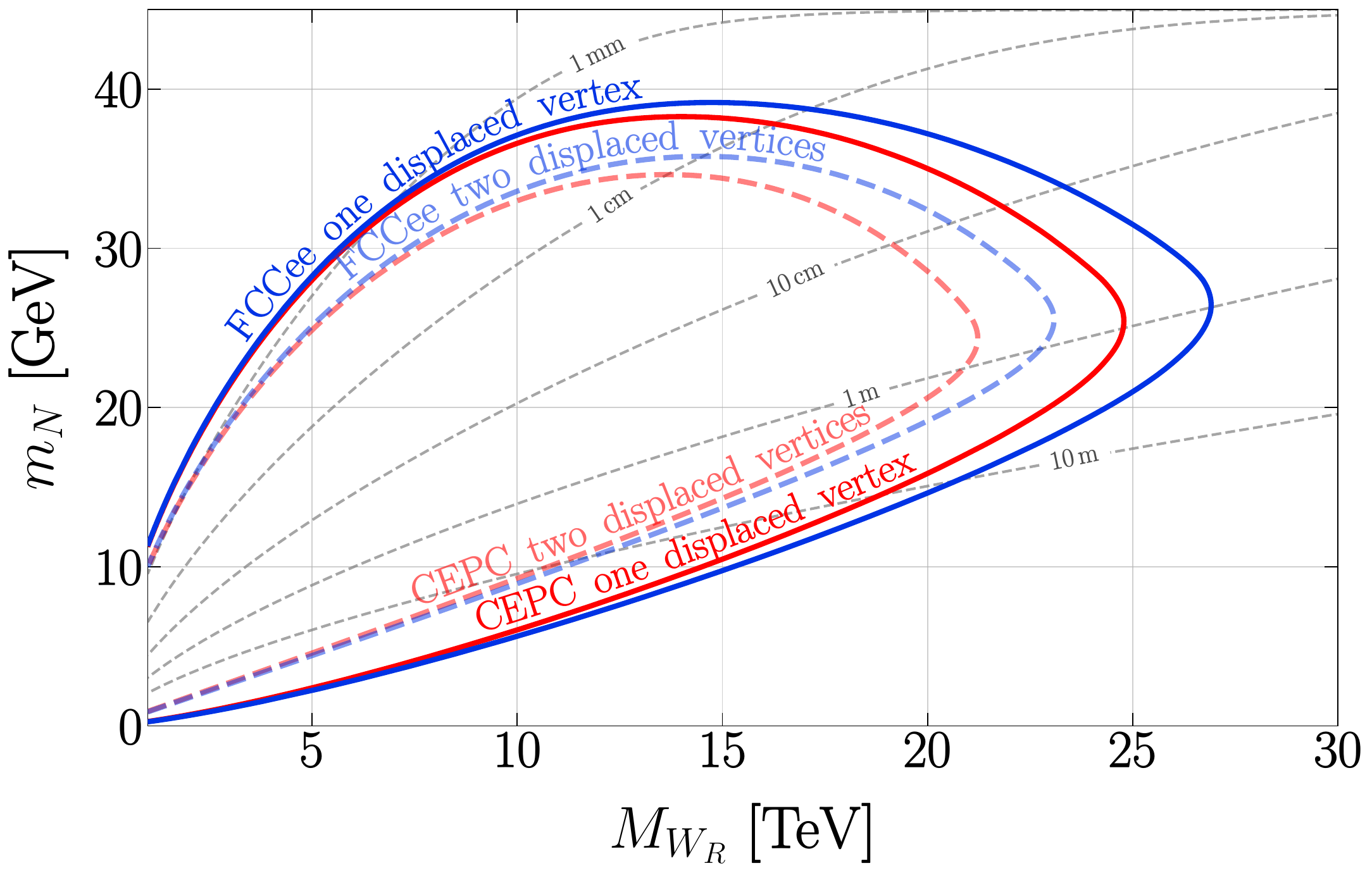}
    \end{minipage}%
    \begin{minipage}{0.45\textwidth}
        \includegraphics[width = \linewidth]{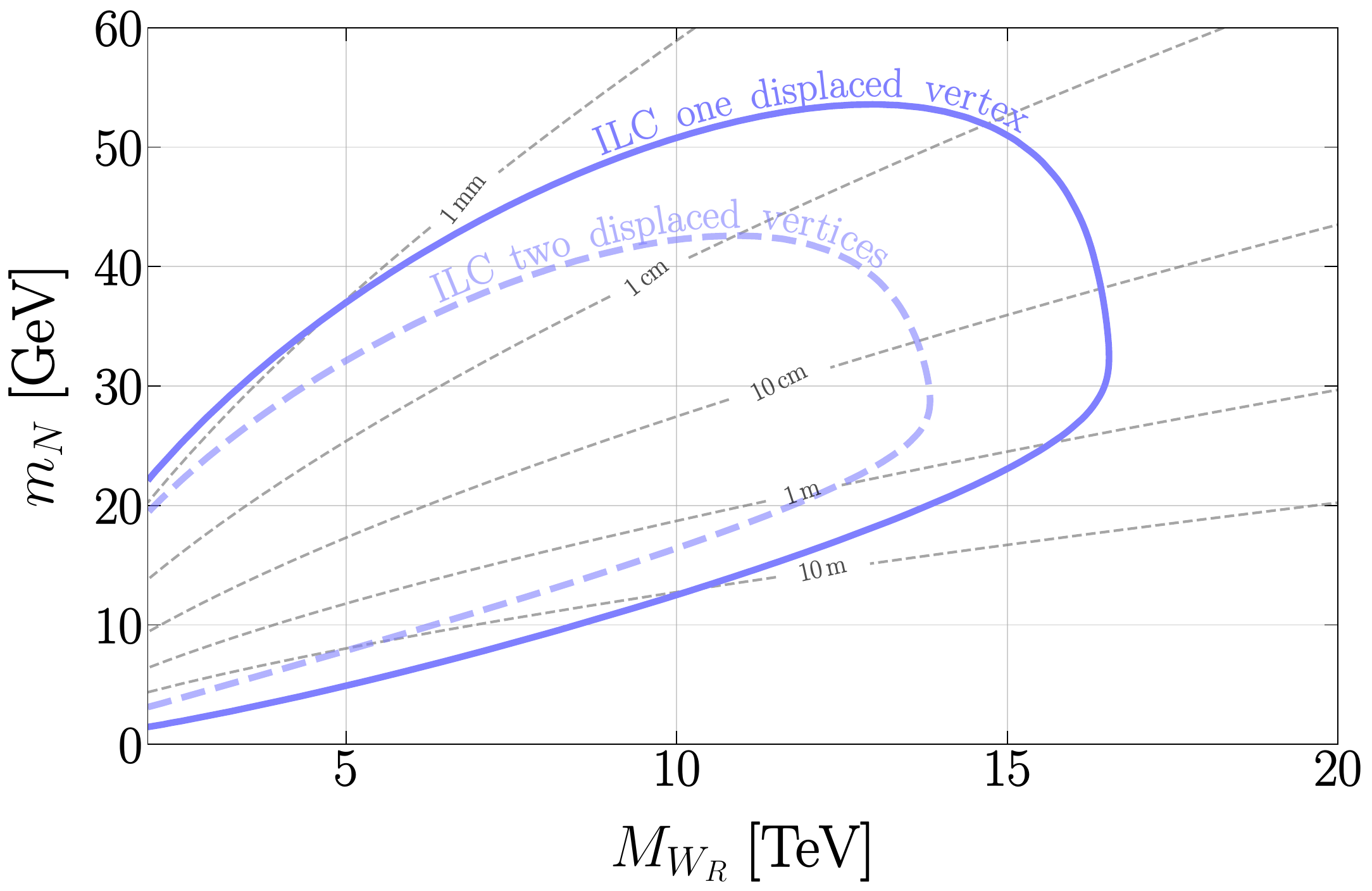}
    \end{minipage}
    \begin{minipage}{0.45\textwidth}
        \includegraphics[width = \linewidth]{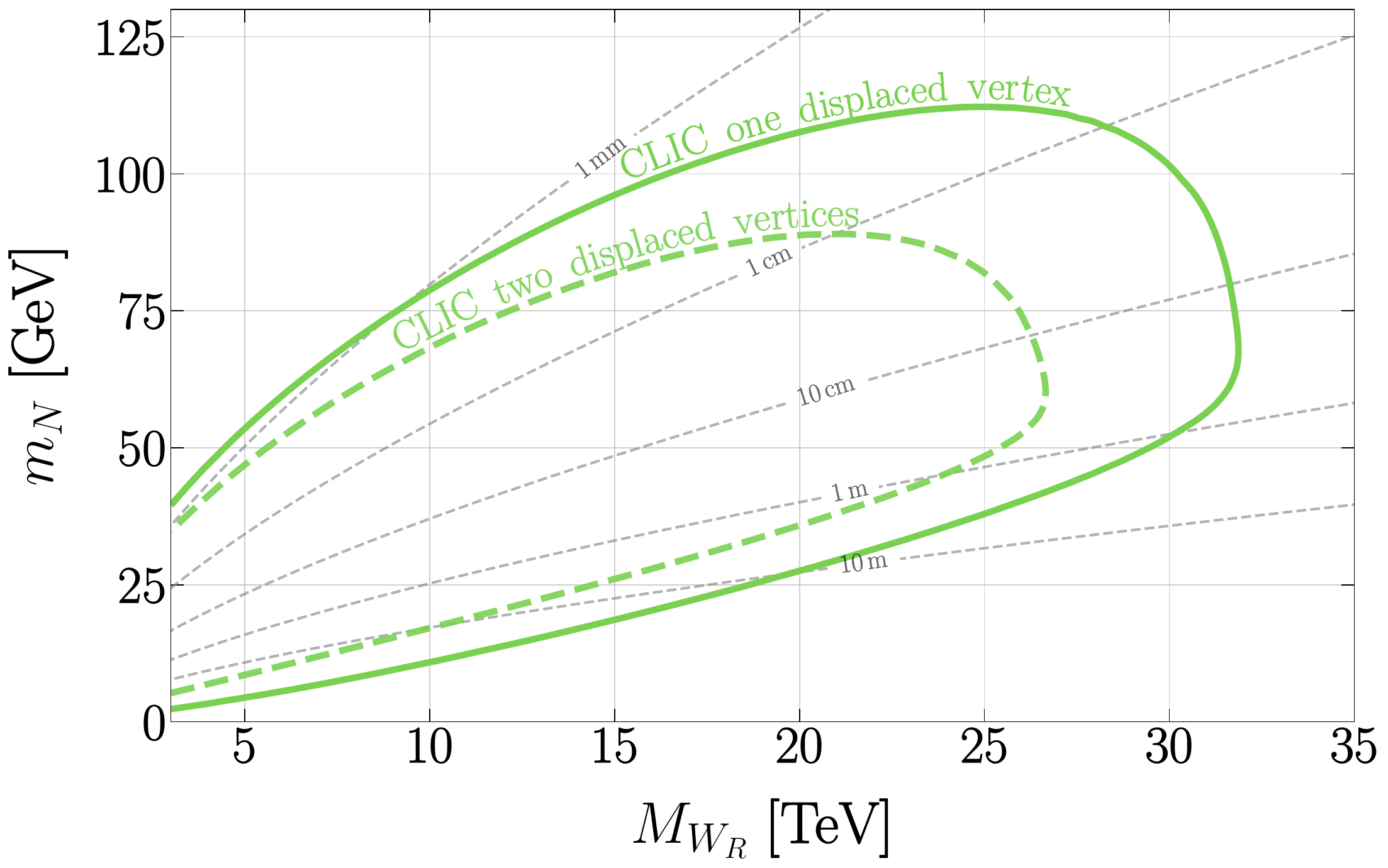}
    \end{minipage}%
    \begin{minipage}{0.45\textwidth}
        \includegraphics[width = \linewidth]{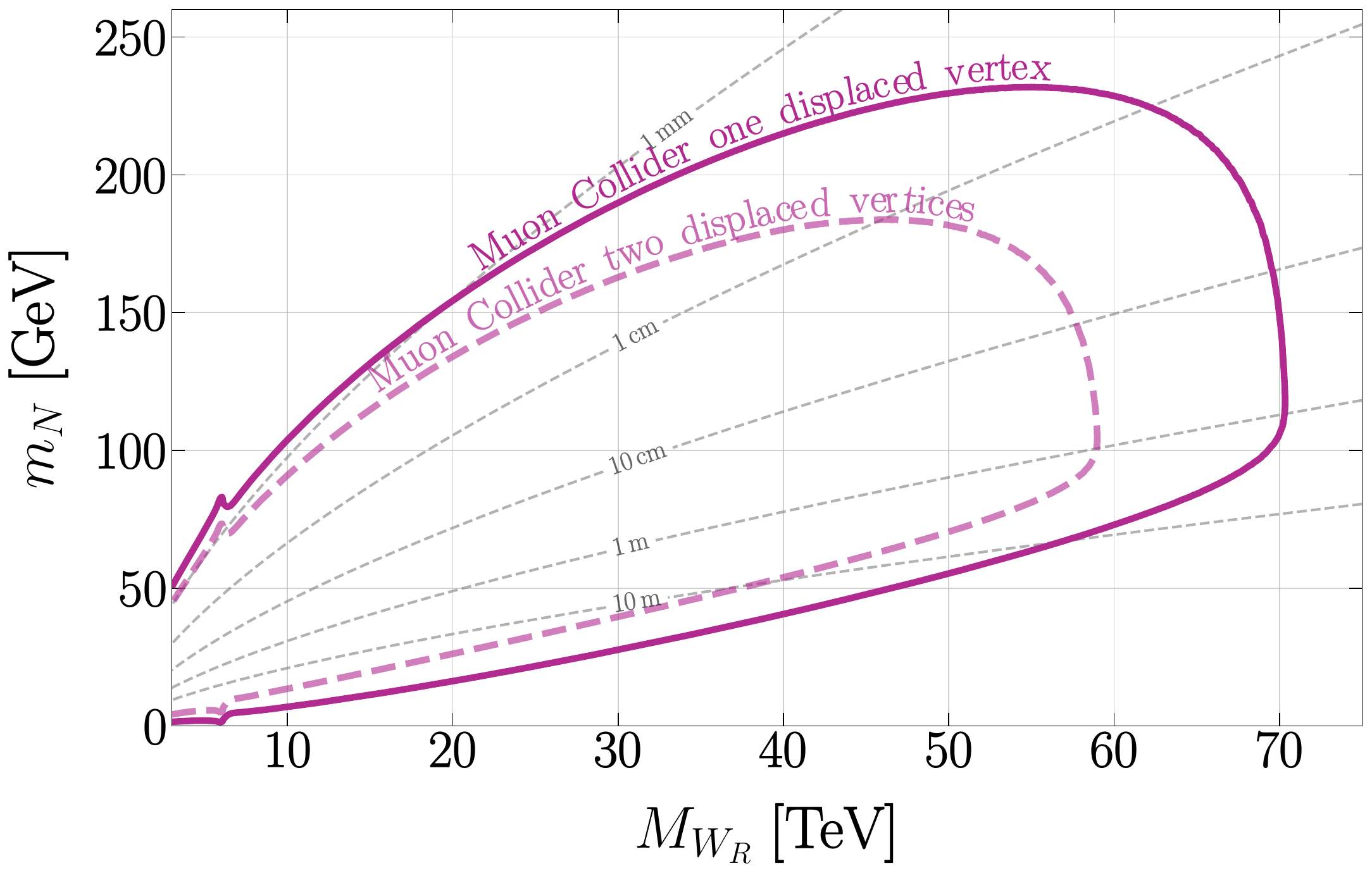}
    \end{minipage}%
    \caption{Sensitivity plots reach using the benchmark points presented in Tables~\ref{tab:energies_luminosities} (FCC-ee and CEPC at the $Z$ pole, and linear and muon colliders the values with the highest center-of-mass energies)
    with a minimum displaced length of \si{5}{mm}; and a maximum displaced length until the tracker volume for FCC-ee, CEPC, ILC and CLIC, and until the muon system for a muon collider (all lengths are shown in Table~\ref{tab:detector_dimensions}).
    The plots were made considering
    $\mathcal{N_{\mathrm{events}}} = 3$, corresponding to a $\sim 95\%$ exclusion limit.}
    \label{fig:sensitivities}
\end{figure*}
The production channels we have explored, together with the decay channels will generate a two lepton and four jets signal ($\ell\ell jjjj$). 50\% of the lepton pair generated are same-sign leptons (SS); this signal is a lepton-number violating signal (LNV) that occurs due to the Majorana nature of HNLs. A diagram for such a process is shown in Fig.~\ref{fig:LNV_diagram}. 

This process has no irreducible SM background, but there are some SM processes that could be a source of background. A preliminary study in the context of a $\mathrm{U}(1)_{B-L}$ SM extension highlighted the decay of heavy hadrons as a potential source of background for this process \cite{Nakajima:2022pkd}. A full background analysis is beyond the scope of this paper, however, we made an estimate of the possible sensitivity that an LNV signal would have in Appendix~\ref{app:prompt_searches}, assuming the optimistic scenario of no background.

As we mentioned previously, given a light $N$ and a sufficiently heavy $W_R$, we could have our HNLs being long-lived and thus generating a displaced vertex (DV). DVs have the advantage of having essentially no SM background (see Fig.~\ref{fig:decay_length} for a comparison between the decay length of $N$, and the decay length of other SM particles, each on their own frame).

Given how we are producing two $N$'s, we would be expecting to produce two displaced vertices. Searches could be done for one DV or for two DVs. Both of these searches have their own set of advantages and disadvantages. 

Searches for one DV will be more sensible to a bigger region of the parameter space, but there might be some ambiguity due to the fact that other models with HNLs also predict single-displaced HNLs, like the minimal type-I seesaw. There could be two ways of removing the ambiguity, we can reconstruct the mass of $N$ and if $m_N > \sqrt{s}/2$ we could be mostly certain that it came from the production of a single HNL, as we would be expecting that the production of an off shell and an on shell HNL to be suppressed. We can also measure the branching ratios of $N$, as different models give different predictions for them, LRSM predicts $N$ to decay mostly semileptonically (unless the $W-W_R$ mixing is nonzero, see Appendix~\ref{app:decay_beta}), whereas the minimal type-I seesaw allows them to decay both leptonically and semileptonically.

Searches for two DV will manage to falsify models that only predict a single DV and would be a clearer indication that what we are seeing is a signal from LRSM,\footnote{LRSM is not the only model which would predict two DV from HNLs. The aforementioned $U(1)_{B-L}$ extension is one of them. Moreover, models that predict the pair production of HNLs naturally appear in different dimension 5 and 6 operators \cite{Mitra:2016kov, Barducci:2022hll}.} but the reconstruction efficiency of two DVs may kill most of the signal. Given how we know that reconstruction efficiency for DVs at LHC detectors decreases linearly with length \cite{Alimena:2019zri, Knapen:2022afb}, the naive assumption would be that for two DVs the reconstruction efficiency would decrease quadratically.

A full analysis regarding the reconstruction efficiency is beyond the scope of this paper. It can depend heavily on the capacities of the proposed detectors, which are nowadays unknown to us. We will thus only work in the optimistic scenario where the reconstruction efficiency is one, as previous studies have done \cite{Blondel:2014bra, Antusch:2016vyf, Antusch:2016ejd, Barducci:2022hll}.
 
\section{Projected sensitivities of displaced searches} \label{sec:results}
We obtain the projected sensitivity for both one DV and two DVs for all colliders mentioned in Sec.~\ref{sec:colliders}. The sensitivity curves are defined by $\mathcal{N}_\mathrm{events} \simeq 3$, corresponding to a $\sim 95\%$ exclusion limit in the limit where there is no background. The number of events for one DV is given by
\begin{equation}
    \label{eq:number_events_1_DV}
    \mathcal{N}_{\mathrm{events}}^{1-\mathrm{DV}} = 
        2\,\mathcal{L}\cdot \sigma\left(\ell_\alpha^- \ell_\alpha^+ \to NN \right) \cdot P_{\mathrm{dec.}}
        \cdot \mathrm{BR}_{\mathrm{vis.}} \cdot \epsilon^{1-\mathrm{DV}}\,,
\end{equation}
and for two DVs, it is given by
\begin{equation}
    \label{eq:number_events_2_DV}
    \mathcal{N}_{\mathrm{events}}^{2-\mathrm{DV}} = \mathcal{L}\cdot \sigma\left(\ell_\alpha^- \ell_\alpha^+ \to NN \right)\cdot P_{\mathrm{dec.}}^2\cdot \mathrm{BR}_{\mathrm{vis.}}^2 \cdot \epsilon^{2-\mathrm{DV}}\,,
\end{equation}
where $\mathcal{L}$ is the integrated luminosity, $\sigma\left(\ell_\alpha^- \ell_\alpha^+ \to NN \right)$ is the production cross section, $P_{\mathrm{dec.}}$ is the probability that the HNLs will be sufficiently displaced and that it will decay within the fiducial volume. $\mathrm{BR}_{\mathrm{vis.}}$ is the visible branching ratio in the fiducial volume, and $\epsilon$ is the detection efficiency. The decay probability is given by:
\begin{equation}
    \label{eq:decay_probability}
    P_{\mathrm{dec.}} = \frac{1}{\pi}\int_{0}^{\pi} \dd \theta\left[ e^{-L_{\mathrm{min}}/L_N} - e^{-L_{\mathrm{max.}}(\theta)/L_N} \right]\,,
\end{equation}
where $\theta$ is the angle with respect to the beam axis, $L_N$ is the decay length of the HNL, which is
\begin{equation}
    L_N = c\,\gamma_N \tau_N = c\,\frac{\sqrt{s - 4m_N^2}}{2\,m_N \Gamma_N}\,.
\end{equation}
$L_{\mathrm{min}}$ is the minimum length for the process to be considered as a displaced signal, which we will take to be around \SI{5}{mm}. \SI{5}{mm} is more than enough to suppress the background from mesons produced in the collider \cite{Alimena:2019zri, Knapen:2022afb}. Even for an $N$ with the mass of the order of a $B$ or a $B_c$, we would still expect the decay length of $N$ to be much higher than mesons (see Fig.~\ref{fig:decay_length}). We can avoid any potential background from heavy mesons with a cut in either displacement or in invariant mass.

\begin{figure*}[t]
    \centering
    \includegraphics[width = 0.5\linewidth]{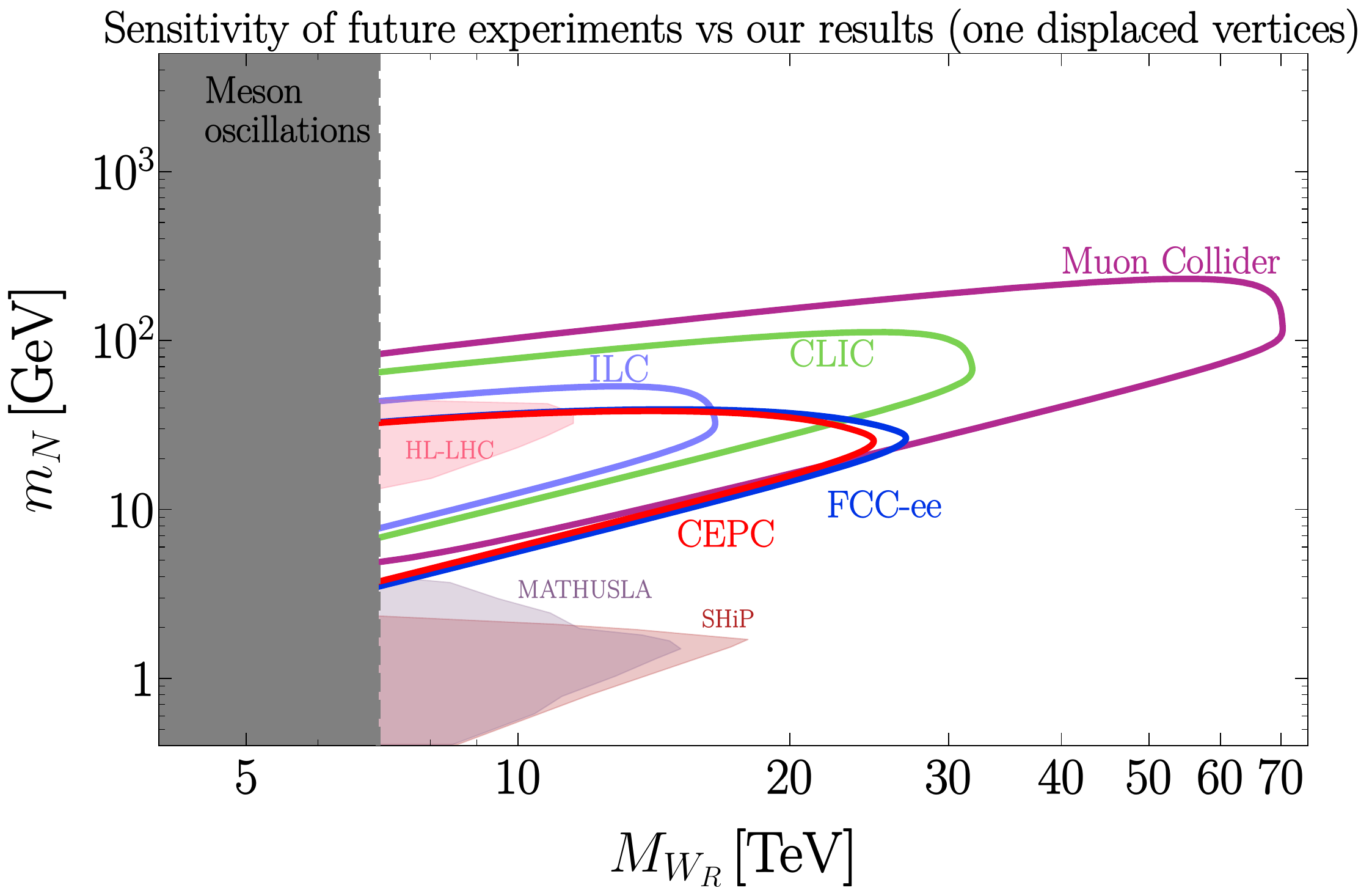}%
    \includegraphics[width = 0.5\linewidth]{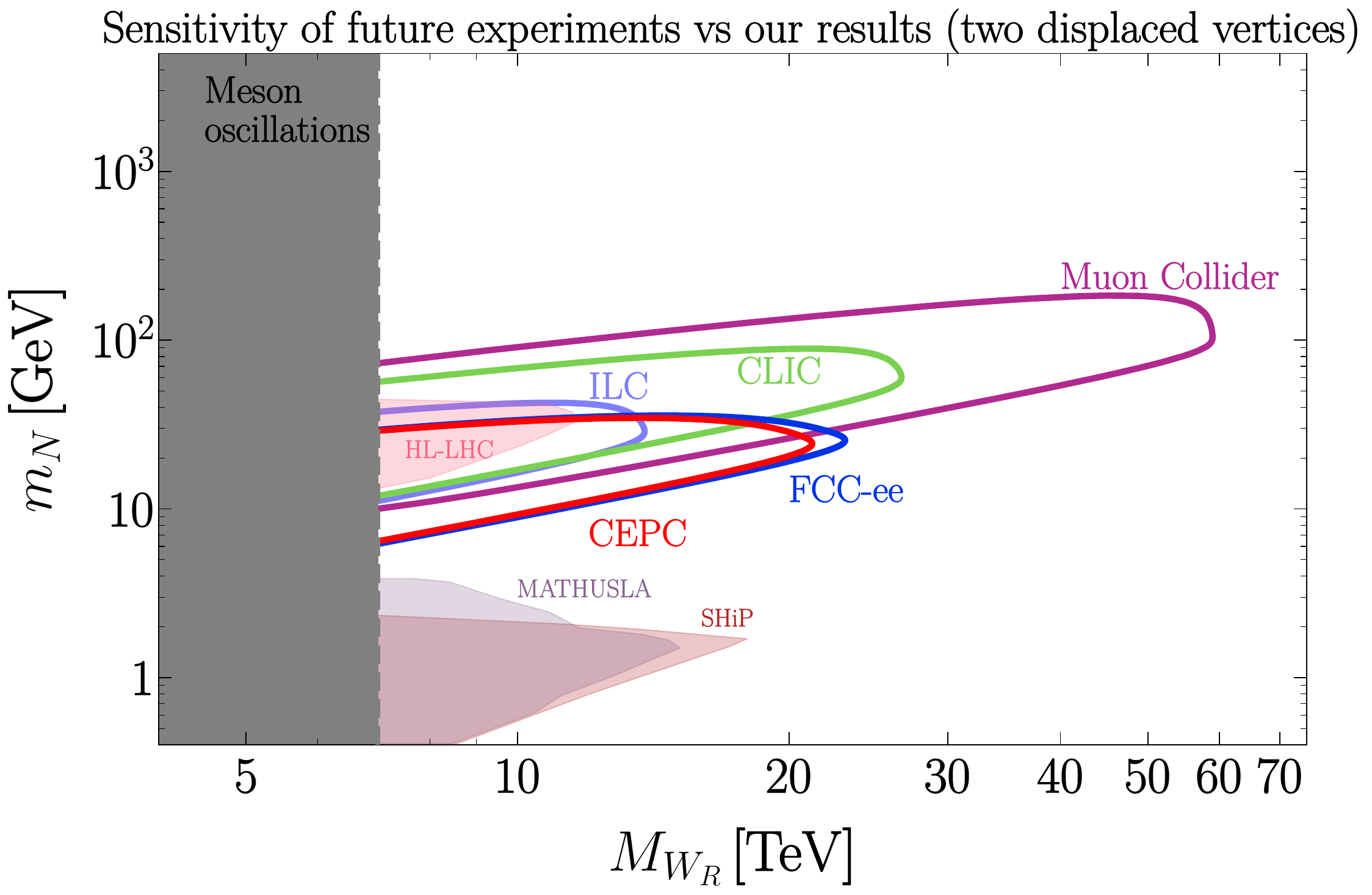}%
    \caption{Comparison between our results shown in Fig.~\ref{fig:sensitivities} and other constraints shown in Fig.~\ref{fig:constraints}. The benchmark points used were highlighted in Fig.~\ref{fig:sensitivities}.}
    \label{fig:comparison_our_results}
\end{figure*}

$L_{\mathrm{max.}}(\theta)$ is the maximum length that the HNLs can have such that they decay within the fiducial volume. $L_{\mathrm{max.}}(\theta)$ will depend on the dimensions of the detector and on $\theta$ due to the cylindrical geometry of the detectors. To a good approximation, we can use the average value of $L_{\mathrm{max.}}(\theta)$ in Eq.\eqref{eq:decay_probability}, which is given by:
\begin{equation}
    \begin{aligned}
    \expval{L_{\mathrm{max.}}(\theta)} = \frac{2}{\pi}\left[\frac{z}{2} \right. &\log(\frac{2R + \sqrt{z^2 + 4 R^2}}{z}) \\
    & + \left.R \log(\frac{z + \sqrt{z^2 + 4 R^2}}{2R}) \right]\,,
    \end{aligned}
\end{equation}
where $R$ and $z$ are the radius and the longitudinal length of the section of the detector in question. Which section of the collider we consider depends on whether the final lepton is an electron or a muon. For muon, we can consider the entire volume of the detector, including the muon system; for electrons, we will only consider the volume until the tracker because electrons that are more displaced can be misidentified to be a different particle. For FCC-ee, CEPC, and the muon collider we will consider the geometry of the IDEA detector; for the linear colliders, we will consider the geometry of the SiD detector.
Table~\ref{tab:detector_dimensions} presents the geometric parameters of different detectors for both their trackers and muon systems.

\begin{table}[ht]
    \centering
    \begin{tabular}{|cl|c|c|}
        \hline
        \multicolumn{2}{|c|}{Detector} & $z/R$ & $\expval{L_{\mathrm{max.}}(\theta)}$  \\ 
        \hline
        \multirow{2}{*}{\shortstack[c]{IDEA \\ (FCC-ee, CEPC)}} & Tracker & $\SI{4.0}{m}/\SI{2.0}{m}$   & \SI{2.2}{m} \\
        & Muon system & $\SI{13.0}{m}/\SI{5.0}{m}$   & \SI{6.4}{m} \\ \hline         
        \multirow{2}{*}{\shortstack[c]{SiD \\ (ILC, CLIC)}} & Tracker & $\SI{1.6}{m}/\SI{1.2}{m}$    & \SI{1.1}{m} \\
         &  Muon system & $\SI{11.3}{m}/\SI{6.0}{m}$    & \SI{6.53}{m} \\
        \hline
    \end{tabular}
    \caption{Dimensions of the proposed detectors for different experiments. The IDEA detector was proposed to both the FCC-ee and CEPC, we took its dimensions from \cite{FCC:2018evy}. The SiD detector was suggested to both the ILC and CLIC and its dimensions were taken from \cite{Bambade:2019fyw}. A muon collider does not yet have a proposed detector, as a benchmark for it we will use the dimensions of the IDEA detector.}
    \label{tab:detector_dimensions}
\end{table}

Given how we are going to work in the simplified scenarios, where $N$ only mixes with $e$ (for $ee$ colliders) and $N$ only mixes with $\mu$ (for the muon collider), we will only be dealing with one length and one specific branching ratio per collider: in both cases we will have that $\mathrm{Br}_{\mathrm{vis}.} = 1$. Different scenarios are treated in Appendix~\ref{app:different_mixing}, where the branching ratio will be affected, how much $N$ can be displaced depending on each channel, and the production.

As we highlighted before, we will work in the optimistic case where the reconstruction (as well as detection) efficiency to be equal to one. This may not be the case for the detection and reconstruction of a pair of displaced vertices, and requires further study.

The cross section $\sigma(\ell_\alpha^+ \ell_\alpha^- \to NN)$ was calculated with the event generator \texttt{WHIZARD 3.0.3} \cite{Kilian:2007gr} and the UFO files from \cite{Roitgrund:2014zka}.\footnote{\texttt{WHIZARD} presented a problem for the computation of cross sections when $\sqrt{s} > M_{Z^\prime}$, this is because the UFO in \cite{Roitgrund:2014zka}. In order to correct for it we needed to include the decay width of $Z^\prime$. Its calculation is shown in Appendix~\ref{app:decay_width_Z}.} The cross section includes initial state radiation (ISR) effects for the FCC-ee, CEPC, and muon collider. For linear colliders we also took into account the effects of beam polarization as described in \cite{Barklow:2015tja} for the ILC and \cite{Roloff:2018dqu} for CLIC.

The projected sensitivities for one and two DVs for each experiment are presented in Fig.~\ref{fig:sensitivities} and in Fig.~\ref{fig:comparison_our_results} in comparison with the sensitivities of future searches.

Both the FCC-ee and CEPC manage to reach very high values of $M_{W_R}$ due to the huge luminosity that these colliders could achieve at the $Z$ pole, thus probing very small values of $\zeta_Z$.
As we stated earlier, the cross section increases quadratically with $\sqrt{s}$. This fact allows linear and muon colliders that have higher values of center-of-mass energy to also probe very high scales of LRSM.

\section{Conclusions} \label{sec:conclusions}
In this paper, we provided a first analysis of the sensitivity that proposed future lepton colliders may have on LRSM parameters for displaced signals of HNLs. 

We first reviewed the relevant theoretical framework, the constraints that stem from different experiments and theoretical contexts, the proposed collider experiments, and the relevant phenomenology of HNLs regarding their production and decay in lepton colliders. We made a particular emphasis on the potential displaced vertex signals that the model could predict given the proper values of $m_N$ and $M_{W_R}$.

We showed that lepton colliders can probe scales of LRSM that no other current experiment could potentially reach. Incidentally, the region of the parameter space that lepton colliders can probe allows for a DM candidate, as well as baryogenesis and the potential lack of CP violation in strong interactions.

Our results should be seen as complementary for searches that could be done at future hadron-hadron colliders like the FCC-hh \cite{Nemevsek:2023hwx}. Given the center-of-mass energy of the proposed lepton colliders, the production of a $W_R$ is impossible. We can, however, reconstruct the mass of $W_R$ either from displaced signals from the lifetime of $N$ and produce it at a hadron-hadron facility.

More detailed studies are yet to be made on the potential that prompt signals may have, and an accurate assessment of the reconstruction efficiency of both single and doubly displaced vertices at future lepton collider facilities.

\section*{Acknowledgements}
I want to particularly thank Oleg Ruchayskiy for multiple discussions surrounding this paper, as well as the encouragement to present it as a single author; and to Giovanna Cottin whose comments greatly improved the quality and presentation of the paper. I would also like to thank Mads Frandsen and Mogens Dam for the comments regarding my master thesis, Juliette Alimena for allowing me to present some results during one of the \textit{LLP at the FCC-ee} talks, and Inar Timiryasov for helpful discussions. 
This project has received funding from the European Research Council (ERC) under the European Union’s Horizon 2020 research and innovation programme (Prgram No. GA 694896) and from the Carlsberg Foundation (grant agreement CF17-0763).

\appendix
\section{CALCULATION OF \texorpdfstring{$\Gamma_{Z^\prime}$}{Lg}} 
\label{app:decay_width_Z}
In order for event generators to work for all regions of the parameter space, including regions where $M_{Z^\prime} \leq \sqrt{s}$, we need to include a nonzero decay width to $Z^\prime$. The calculation of $\Gamma_{Z^\prime}$ is straightforward: we simply need to know how $Z^\prime$ interacts with the rest of fermions,
\begin{equation}
    \label{eq:Z_interactions}
    \mathcal{L}_{Z^\prime} = \frac{g\,c_{w}}{c_{2w}^{1/2}}\,\bar{\Psi} \slashed{Z}^\prime \left[t_{R}^3\,P_R + t_{w}^2 t_L^3\,P_L - t_{w}^2 q \right]\Psi\,,
\end{equation}
where $\Psi$ is any SM fermion field, $t_L^3$ is the value of the left-handed weak isospin of $\Psi$, $t_R^3$ is the right-handed weak isospin of $\Psi$, and $q$ is the charge of $\Psi$. There are other interactions that arise from the mixing between neutral gauge bosons, much like the second term in Eq.\eqref{eq:Z_interactions_N}, but these sorts of interactions are very suppressed and should not be necessary for a first-order approximation for $\Gamma_{Z^\prime}$. 

We can rewrite Eq.\eqref{eq:Z_interactions} as
\begin{equation}
    \mathcal{L}_{Z^\prime} = \frac{g\,c_{w}}{c_{2w}^{1/2}}\, \bar{\Psi} \slashed{Z}^\prime \left[ g_R P_R + g_L P_L \right] \Psi\,,
\end{equation}
where $g_L$ and $g_R$ depend on the fermion in question and its different quantum numbers, Table~\ref{tab:gL_gR} depicts their values for all SM fermions. We can then get the decay width to an arbitrary pair of massless fermions,
\begin{equation}
    \Gamma(Z^\prime \to \Psi \bar{\Psi}) = \frac{g^2}{24\pi} \frac{c_w^2}{c_{2w}} M_{Z^\prime}\left(g_L^2 + g_R^2 \right)\,,
\end{equation}
then we can get the total decay width as the sum over all fermion fields,
\begin{equation}
    \Gamma_{Z^\prime} = \sum_{\Psi} N_{\Psi} \Gamma(Z\to \Psi \bar{\Psi})\,,
\end{equation}
where $N_{\Psi}$ is a number to account for the different quark colors and identical particles for Majorana particles, it is 3 for quarks, 1 for charged leptons, and $1/2$ neutral leptons. 

\begin{table}[ht]
    \begin{tblr}{b{2.5cm} b{2.5cm}>{\centering} b{2.5cm}>{\centering}}
        \hline
        \hline
        Particle type & $g_L$ & $g_R$ \\ \hline
         $U$ & $-\frac{1}{6}t_w^2$ & $\frac{1}{2} - \frac{2}{3} t_w^2$ \\
         $D$ & $-\frac{1}{6}t_w^2$ & $-\frac{1}{2} + \frac{1}{3} t_w^2$ \\
         $\ell$ & $\frac{1}{2}t_w^2$ & $-\frac{1}{2} + t_w^2$ \\
         $\nu$ & $\frac{1}{2}t_w^2$ & $-\frac{1}{2}t_w^2$ \\
         $N$ & $-\frac{1}{2}$ & $\frac{1}{2}$ \\ \hline
    \end{tblr}
    \caption{Values of $g_L$ and $g_R$ for the different types of particles. $U$ and $D$ are any type of up and down quark respectively, $\ell$ are charged leptons, $\nu$ are active neutrino, and $N$ are HNLs.}
    \label{tab:gL_gR}
\end{table}

\section{CONSIDERING A NONZERO \texorpdfstring{$\beta$}{Lg}}

\begin{figure}
    \includegraphics[width = \linewidth]{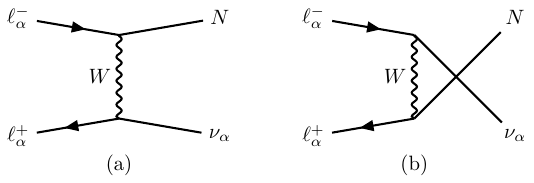}
    \caption{Production diagrams of an HNL and a neutrino.}
    \label{fig:single_production_HNL}
\end{figure}

\begin{figure}[t]
    \includegraphics[width = \linewidth]{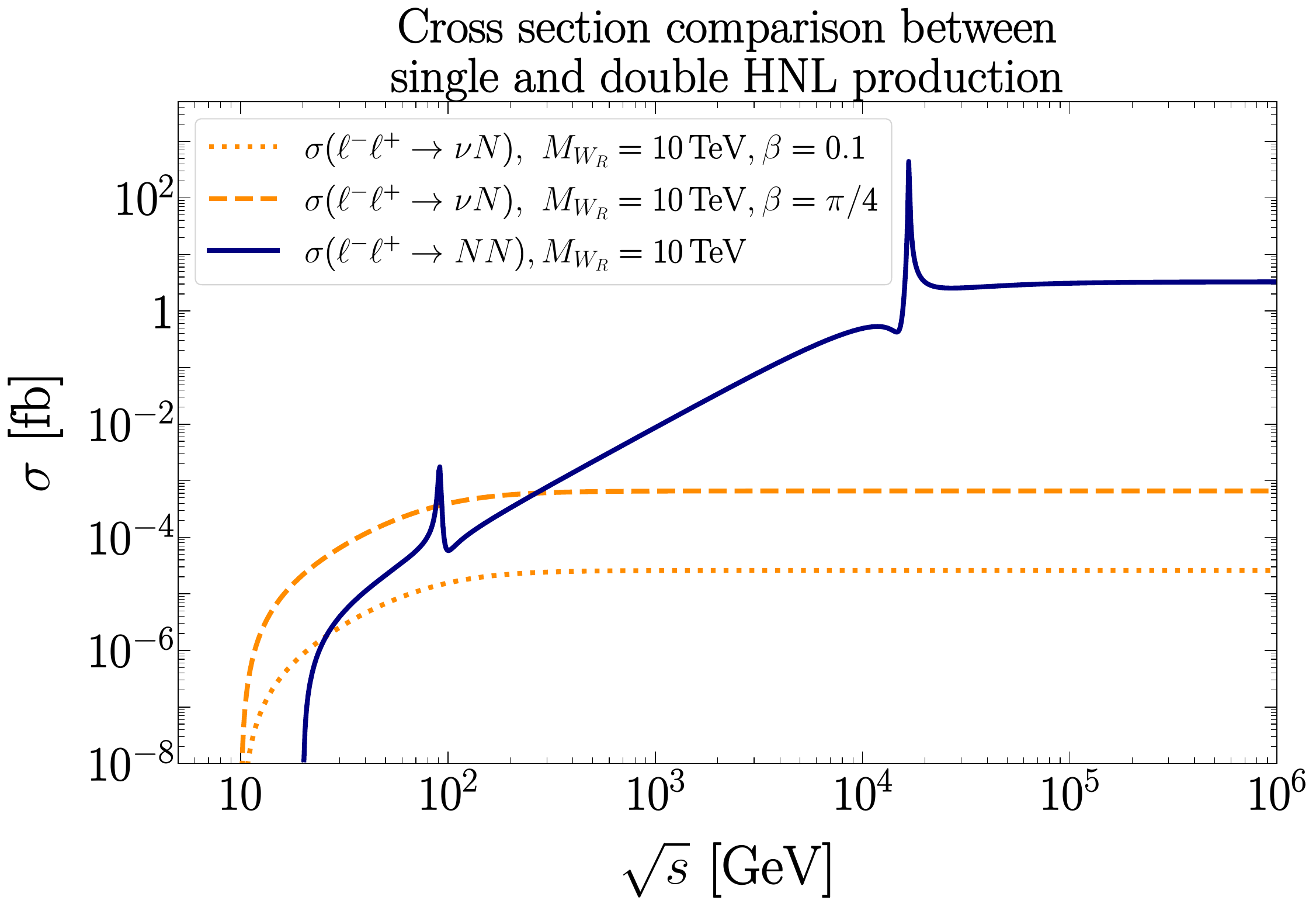}
    \caption{Cross section comparison between the production of a single HNL and the production of two HNLs.}
    \label{fig:single-vs-double-cross section}
\end{figure}

In the main text, we considered the simplified case where $\beta = 0$, the value that dictates the mixing of the vev in the bidoublet in [see Eqs.~\eqref{eq:scalars} and \eqref{eq:vevs}]. If we were not to neglect it, then this would open a new set of interactions for $N$, as we would now have a tree-level interaction with the SM $W$ field from the mixing angle $\xi_W$, [see Eqs.~\eqref{eq:W_interactions_N} and \eqref{eq:mixing_W}]. That could affect both the decay and production of HNLs.

The allowed values of $\beta$ or $\alpha$ depend on whether we're considering $\mathcal{C}$ or $\mathcal{P}$. In the case of $\mathcal{C}$, both values are unrestricted $0 \leq \beta < \pi/4$ and $0 < \alpha \leq 2\pi$. Whereas for the case of $\mathcal{P}$, there is a bound on a specific combination that arises from considerations in quark mass matrices \cite{Maiezza:2010ic, Senjanovic:2014pva, Senjanovic:2015yea},
\begin{equation}
    \sin(\alpha) \tan(2\beta) \lesssim \frac{2 m_b}{m_t}\,,
\end{equation}
where $m_b$ is the mass of the bottom quark, and $m_t$ of the top quark.

To have an idea of the potential that $W_R-W$ mixing could have on our bounds, we will choose two benchmark points $\beta = 0.1, \pi/4$ (none of our results will depend on the specific value of $\alpha$, so we will not assume any value for it). We should stress, that the value $\beta = \pi/4$ is unphysical, as it would predict up and down type quarks to have the same masses, we will only use it to consider the full potential that the $W - W_R$ mixing would have.

\subsection{Effects at production} \label{app:production_beta}

A nonzero $\xi_W$ allows for the production of a single HNL and an active neutrino through a $W$-boson exchange in a $t$ and $u$ channel; both diagrams are shown in Fig.~\ref{fig:single_production_HNL}. 

As we mentioned before, there are other models that generate the same signal, such as the type-I seesaw. We could only distinguish models that generate such a signal from a measurement of the branching ratio of the decays of the HNLs produced and/or lifetimes.

As we can see from Fig.~\ref{fig:single-vs-double-cross section} the limit where $\beta \to \pi/4$, the production cross section of a single HNL in the LRSM is competitive with the production of two HNLs near the $Z$-pole; however, for higher values of $\sqrt{s}$, the cross section is much smaller than the production of two HNLs. This is because for higher $\sqrt{s} > M_{W}$ then $\sigma(\ell \ell \to N \nu) \propto M_{W}^2/M_{W_R}^4\,\sin^2(2\beta)$, as opposed to the behavior of the cross section of the doubly produced HNLs shown in Eq.~\eqref{eq:cross_section_behaviour}, which grows with $s$ for some $\sqrt{s}$ and then plateaus to $\sigma(\ell \ell \to N N) \propto 1/M_{W_R}^2$. 

For smaller and more realistic values of $\beta$, then the production of a single HNL and a neutrino should be suppressed in most cases, as it is shown in Fig.~\ref{fig:single-vs-double-cross section} for the benchmark point of $\beta = 0.1$. 

A potential discrimination between LRSM and other models that predict single HNL production at the $Z$ pole, like the type-I seesaw may be more difficult in this case, especially since for a nonzero $\beta$ the lifetime and branching ratios may change by quite a lot (the change in lifetime and its effects on our projected sensitivity are discussed in Appendix~\ref{app:decay_beta}). Indeed, a non-zero $\beta$ allows now for charged-current mediated leptonic decays. A more dedicated study might be needed as a way to discriminate between both models.

\subsection{Effects at decay} \label{app:decay_beta}
\begin{figure}
    \centering
    \includegraphics[width = \linewidth]{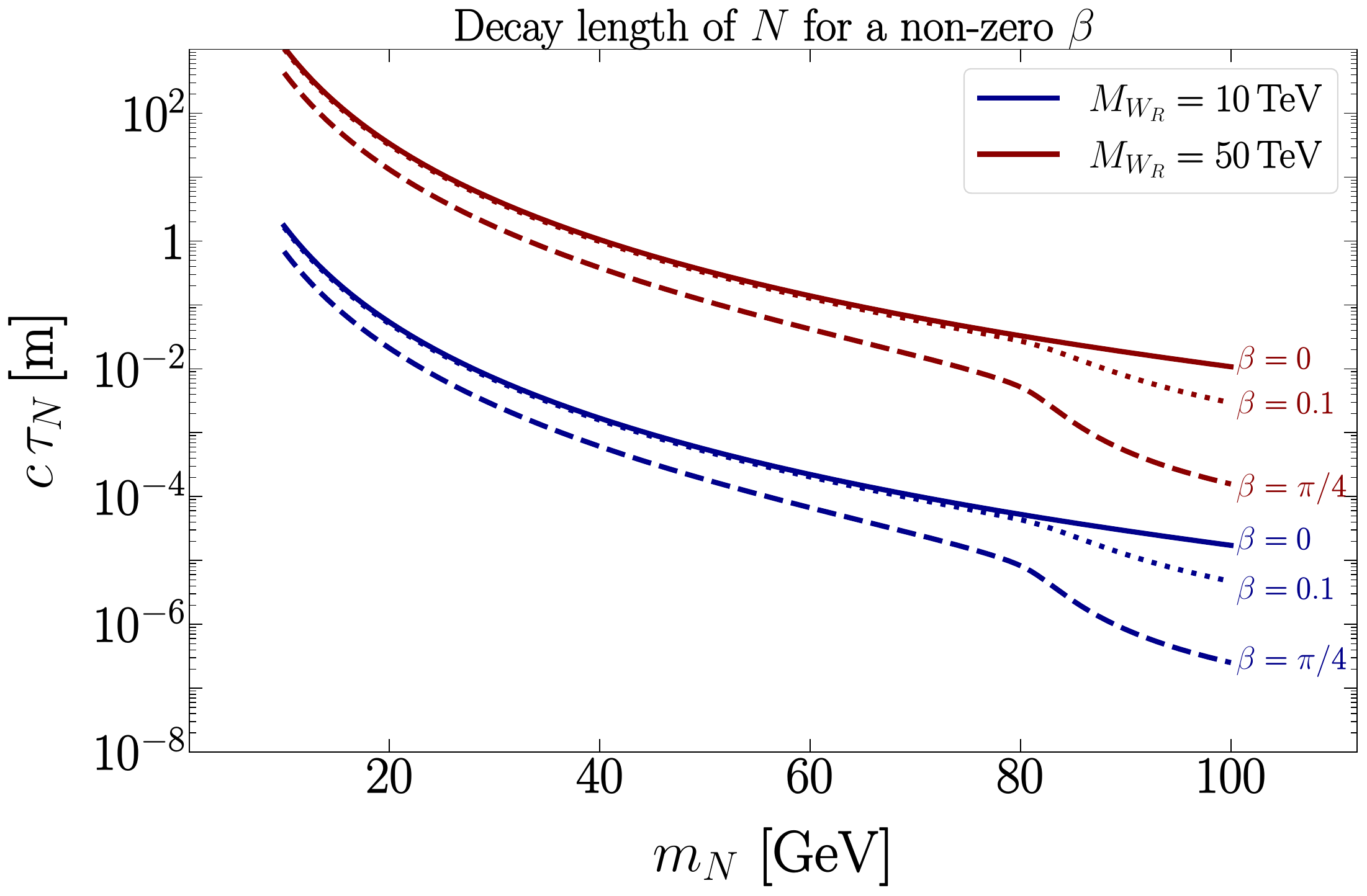}
    \caption{The effect that $\beta$ may have on the lifetime on HNLs for $M_{W_R} = \SI{10}{TeV}$.}
    \label{fig:lifetime_beta}
\end{figure}

\begin{figure*}
    \begin{minipage}{0.45\textwidth}
        \includegraphics[width = \linewidth]{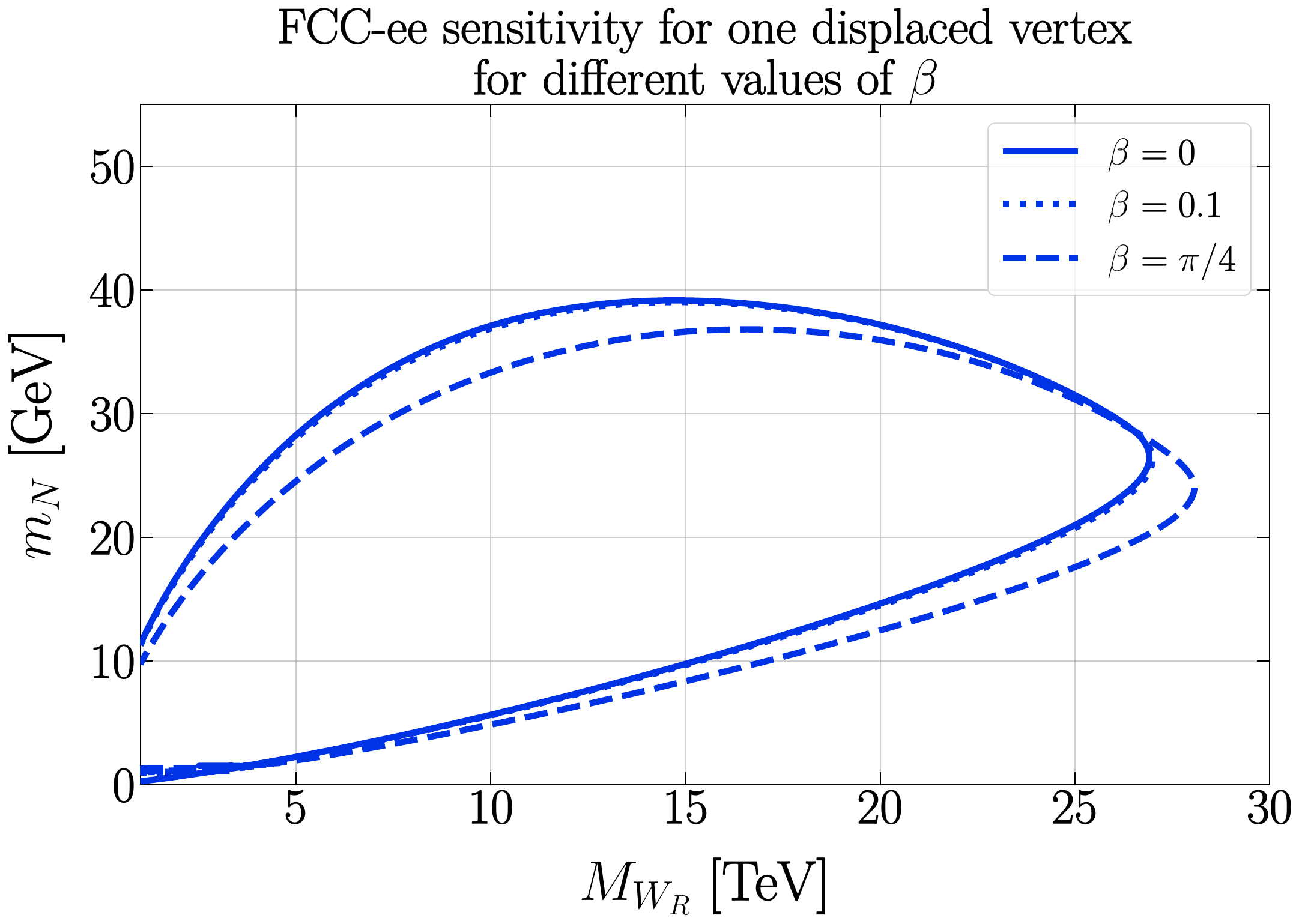} 
    \end{minipage}%
    \begin{minipage}{0.45\textwidth}
        \includegraphics[width = \linewidth]{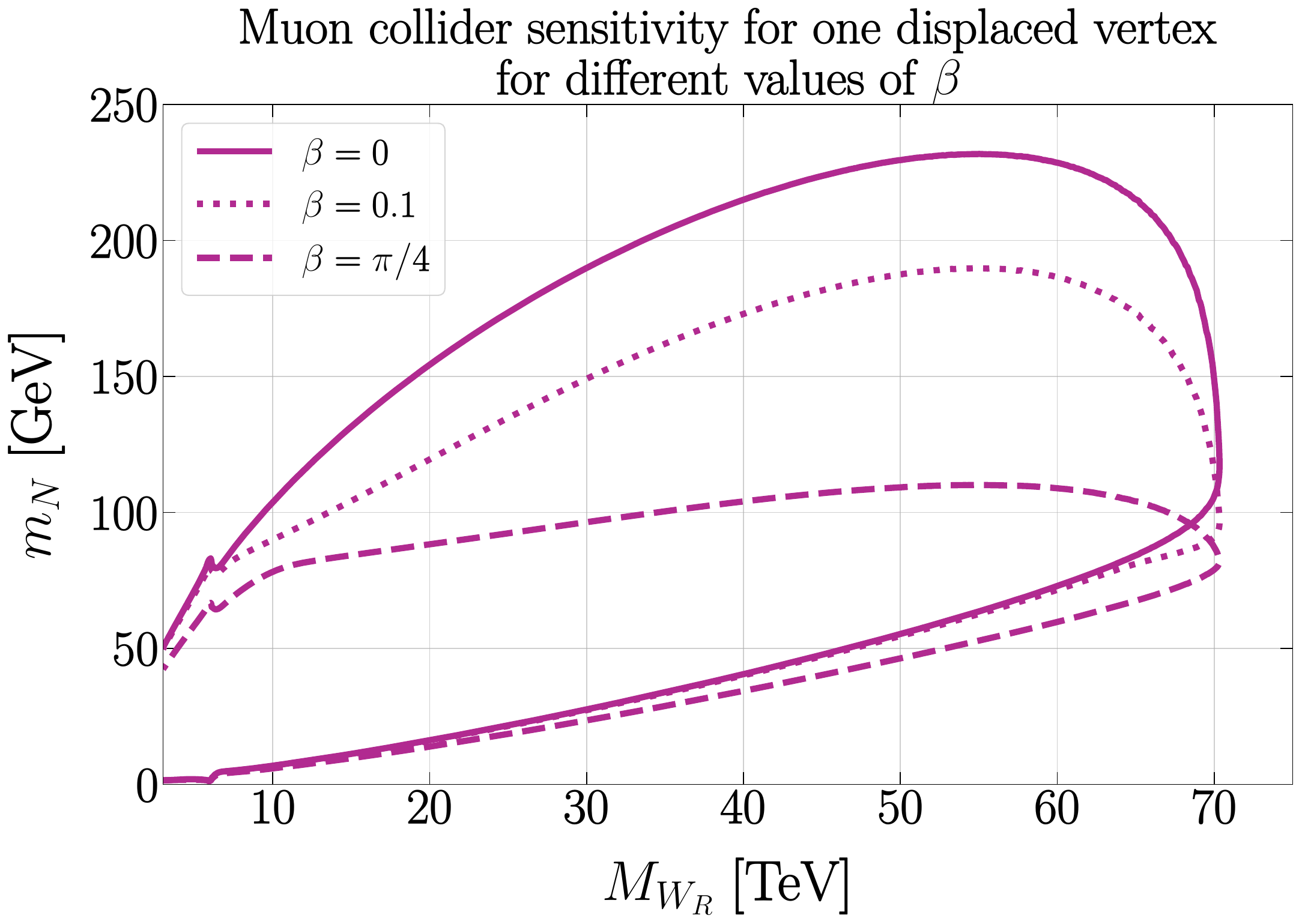} 
    \end{minipage}
    \caption{The difference in sensitivity when considering maximum for different values of $\beta$ for FCC-ee and muon collider for a single DV using the same bechmark parameters as used in Fig.~\ref{fig:sensitivities}.}
    \label{fig:sensitivites_beta}
\end{figure*}
A nonzero value of $\xi_W$ also affects the lifetime of HNLs, since it allows decays mediated through an on shell or off shell $W$ boson. These decay channels will allow $N$ to have leptonic decays ($N \to \ell \ell \nu$) as well as to add more semileptonic decay channels ($N \to \ell q q$). To encapsulate all these effects, we can generalize Eq.~\eqref{eq:decay_width_approximation} to include on and off shell decays to a $W$ boson,  
\begin{equation}
    \Gamma_{\mathrm{tot.}} \simeq \frac{G_F^2 m_N^5}{16\pi^3} \left[\frac{M_W^4}{M_{W_R}^4} + \frac{3}{2}\abs{\xi_W}^2 I(m_N, m_W, \Gamma_W) \right]\,,
\end{equation}
where we have also ignored the masses of all final particles. The second term is the contribution from the $W$-mediated boson decay, the factor $3/2$ comes from the three extra leptonic decay channels. The dimensionless function $I$ comes from a phase space integration and is defined as
\begin{equation}
    I(m_N, M_W, \Gamma_W) = \frac{1}{2}\int_0^1 \dd x \frac{(1-x)^2 (1+2x)}{(1 - x m_N^2/M_W^2)^2 + \Gamma_W^2/M_W^2}\,,
\end{equation}
where $\Gamma_W$ is the decay width of the $W$ boson. 

The addition of these decay channels will directly affect the lifetime of $N$. 
If $m_N > M_W$ then we allow for $N$ to also decay to an on shell $W$ boson and reduce the lifetime of $N$ depending on the value of $\beta$.
For smaller values of $m_N$, the lifetime will still be affected, but in a much less dramatic way. We can see how much the decay length would change for our different benchmark points of $\beta$ in Fig.~\ref{fig:lifetime_beta}. Indeed, as we would expect, for $\beta = 0.1$ then the decay length and lifetime only significantly change for $m_N > M_{W}$. But for $\beta = \pi/4$ where $\xi_W$ is maximum, then $N$ decays mediated by $W$ and $W_R$ are equally likely and lifetime changes for all values of $m_N$, especially for $m_N > M_W$.

The difference in lifetime changes the sensitivities for searches of displaced HNLs. For example, in the case of circular colliders, the change would not be as significant, but for linear colliders and for a muon collider, where the sensitivity reaches values higher than the $W$ mass, the constraints change dramatically, as HNLs would no longer be long-lived.

We plotted how much the projected sensitivity would change for our results from the FCC-ee and muon collider in Fig.~\ref{fig:sensitivites_beta} for the number of single HNL displaced vertices. The projections for the FCC-ee do not change as much for either of our benchmark points of $\beta = 0.1$, whereas for $\beta = \pi/4$, we become more sensible to lighter values of $m_N$ and higher values of $M_{W_R}$. Whereas for a muon collider, our projections significantly change due to the fact that a muon collider will be more sensible to HNLs with masses higher than $M_W$. We would expect similar results for both the ILC and CLIC, as they are also sensible to heavy HNLs.

A nonzero $\beta$ will also change the branching ratios as it will allow for leptonic and semileptonic decays instead of only semileptonic decays. It is still too early to properly know whether the reconstruction efficiencies would change for semileptonic or leptonic signals, but this would affect our naive bounds shown in Fig.~\ref{fig:sensitivites_beta}.

\section{CONSIDERING DIFFERENT MIXING SCENARIOS} \label{app:different_mixing}
\begin{figure*}
    \begin{minipage}{0.45\textwidth}
        \includegraphics[width = \linewidth]{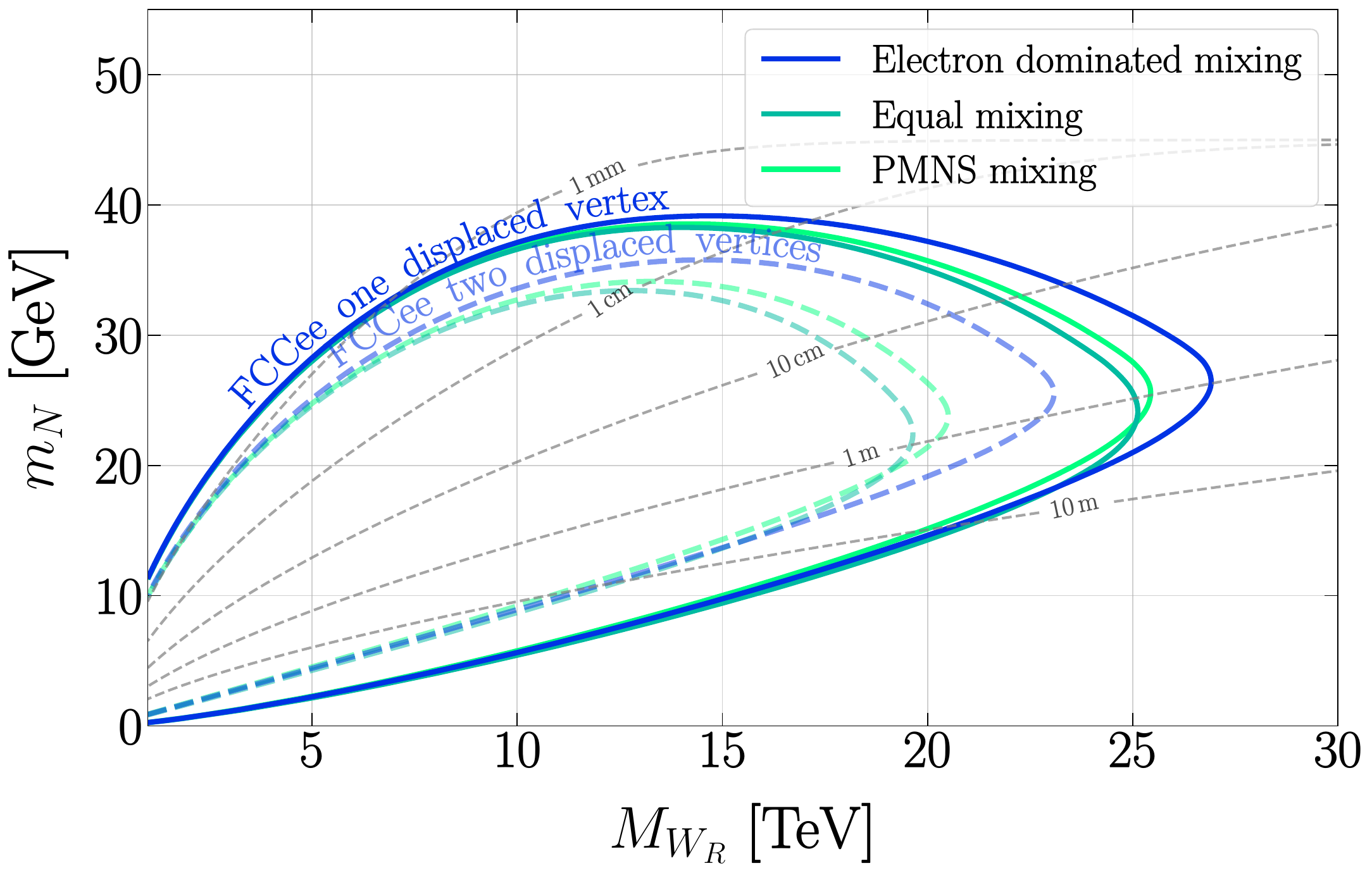}
    \end{minipage}
    \begin{minipage}{0.45\textwidth}
        \includegraphics[width = \linewidth]{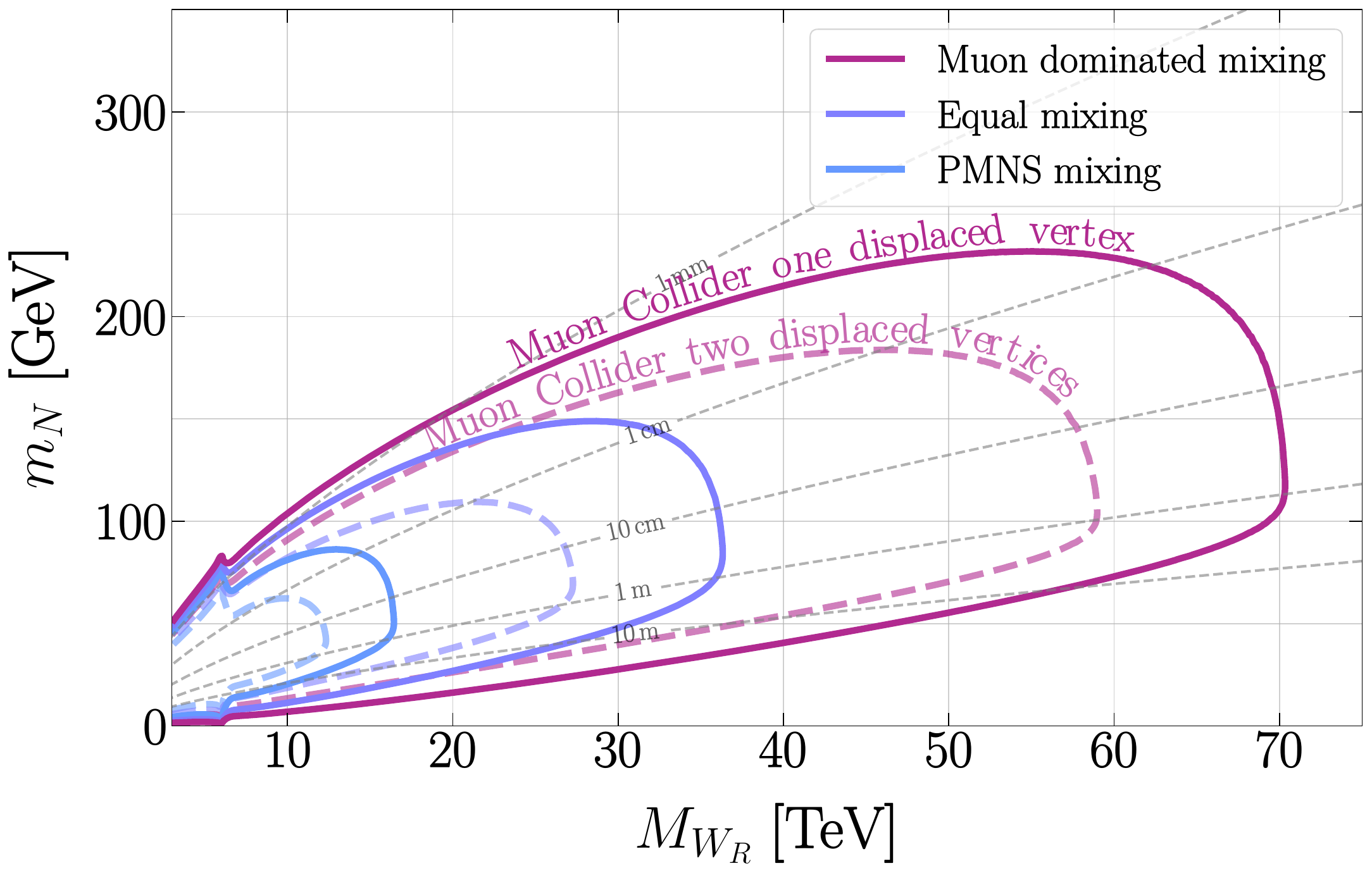}
    \end{minipage}
    \caption{95\% exclusion limit for different mixing scenarios, where we used the same benchmark points as in Fig.~\ref{fig:sensitivities} for both FCC-ee and a muon collider. The electron and muon dominated scenarios are the case where $\abs{U_{e1}}^2 = 1$ and $\abs{U_{\mu 1}}^2 = 1$, the equal mixing case is where $\abs{U_{e1}}^2 = \abs{U_{\mu 1}}^2 = 1/3$, and the PMNS mixing is where $\abs{U_{e1}}^2 = \num{0.68}$ and $\abs{U_{\mu 1}}^2 = \num{0.07}$.} 
    \label{fig:different_mixing_sensitivities}
\end{figure*}

\begin{figure*}
    \begin{minipage}{0.45\textwidth}
        \includegraphics[width = \linewidth]{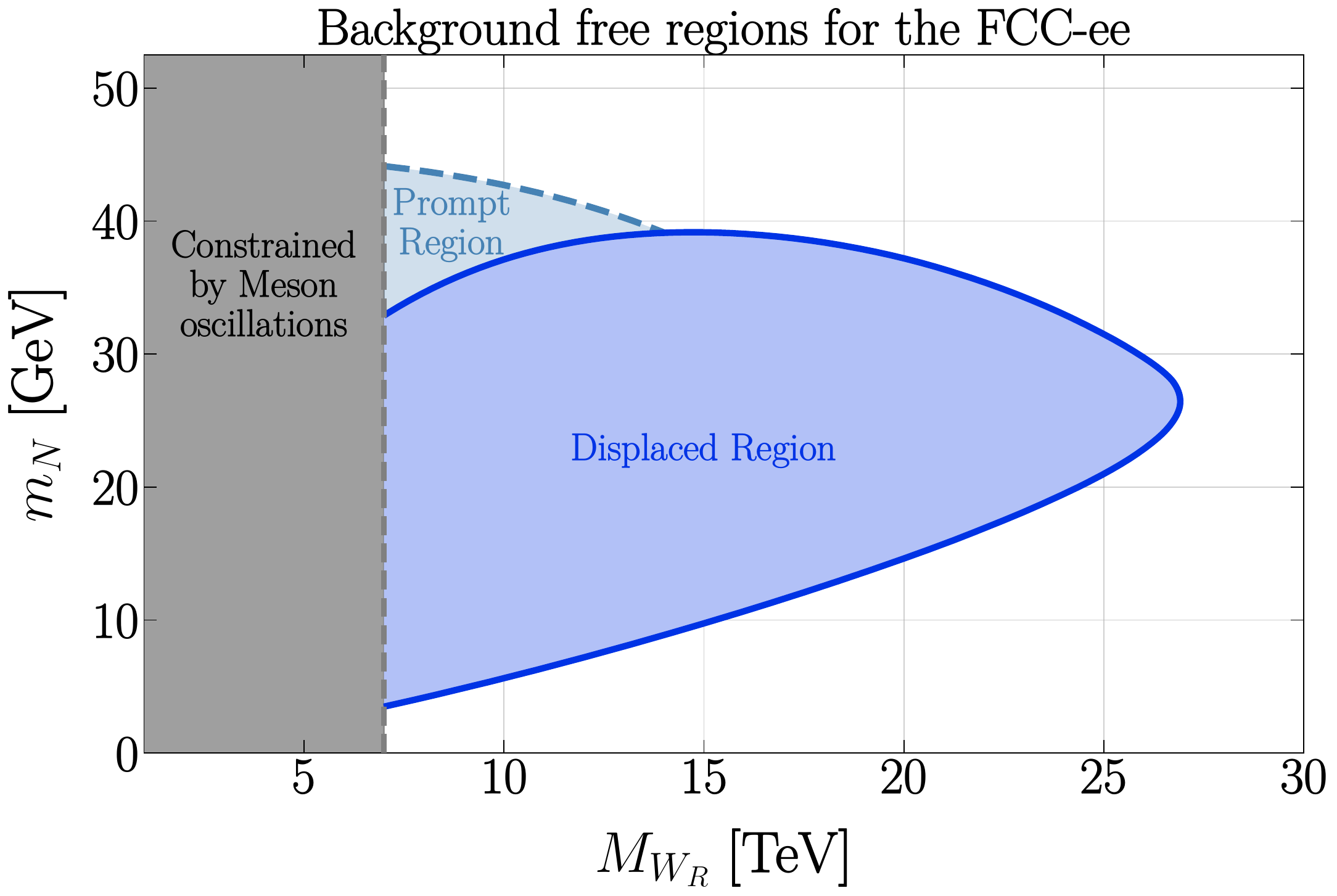} 
    \end{minipage}
    \begin{minipage}{0.45\textwidth}
        \includegraphics[width = \linewidth]{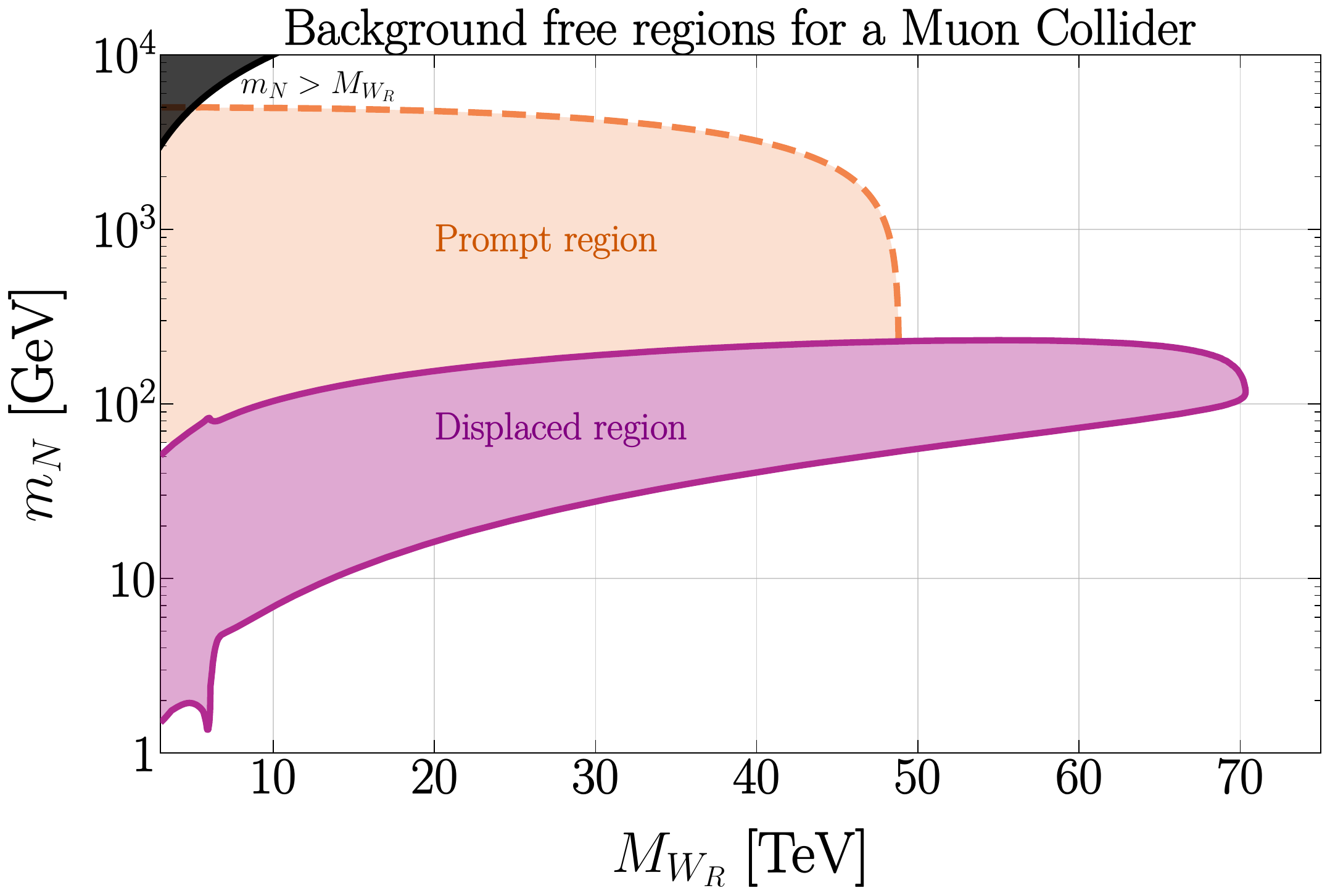} 
    \end{minipage}
    \caption{95\% exclusion limit that prompt LNV would have to future colliders compared to displaced searches. The plot on the left shows the potential that the FCC-ee and the muon collider would have for the same benchmark points considered in Fig.~\ref{fig:sensitivities}. We highlighted the prompt and displaced regions (of one single displaced vertex) of the parameter space. For the plot on the left, we included the area which would have already been constrained from meson oscillations, and on the right the region in which the theory would not be considered perturbative.}
    \label{fig:prompt_sensitivities}
\end{figure*}

In the main text, we only considered the case where $\abs{U_{e1}}^2 = 1$ (for $ee$ colliders) and where $\abs{U_{\mu 1}}^2 = 1$ (for the muon collider). The formulas we have considered in Eqs.~\eqref{eq:number_events_1_DV} and \eqref{eq:number_events_2_DV} are in the simplified scenarios. In a more general case, then we would have
\begin{align}
    \mathcal{N}_{\mathrm{events}}^{1-\mathrm{DV}} &= 
    \begin{aligned}[t]
        &2\cdot \sigma(\ell_\alpha^- \ell_\alpha^+ \to NN) \cdot \\ 
        &\left(P_\mathrm{dec.}^{\mathrm{tracker}}\,\mathrm{BR}_e + P_\mathrm{dec.}^{\mu-\mathrm{sys.}}\,\mathrm{BR}_\mu  \right)\,,
    \end{aligned} \\
    \mathcal{N}_{\mathrm{events}}^{2-\mathrm{DV}} &= 
    \begin{aligned}[t]
        &\sigma(\ell_\alpha^- \ell_\alpha^+ \to NN) \cdot \\ 
        &\left(P_\mathrm{dec.}^{\mathrm{tracker}}\,\mathrm{BR}_e + P_\mathrm{dec.}^{\mu-\mathrm{sys.}}\,\mathrm{BR}_\mu  \right)^2\,,
    \end{aligned}
\end{align}
where $P_\mathrm{dec.}^{\mathrm{tracker}}$ and $P_\mathrm{dec.}^{\mu-\mathrm{sys.}}$ are the decay probabilities until the tracker and muon system respectively (the lengths for the detectors considered are in Table~\ref{tab:detector_dimensions}), $\mathrm{BR}_e$ and $\mathrm{BR}_\mu$ are the branching ratios of $N$ decaying to electrons and muons respectively, and where we assumed that a reconstruction/detection efficiency $\epsilon \simeq 1$ for both single and doubly displaced signals for both electrons and muons. 

Beyond the simplified scenarios, the production cross section will also change, in particular for the $W_R$ mediated $t$ and $u$ channels, as shown in Eq.~\eqref{eq:cross_section_behaviour}, which means that there will be a decrease in sensitivity proportional to $\abs{U_{e1}}^4$ for linear colliders and to $\abs{U_{\mu 1}}^4$ for a muon collider. Moreover, there is also a suppression from the branching ratios, since $\mathrm{BR}_\alpha = \abs{U_{\alpha 1}}^2$.

We can estimate how much sensitivity we would lose between our simplified scenarios and with general mixing, for $P_\mathrm{dec.}^{\mathrm{tracker}} \simeq P_\mathrm{dec.}^{\mu-\mathrm{sys.}}$. For one DV, it would be
\begin{align}
    \frac{\mathcal{N}_{\mathrm{events}}^{1-\mathrm{DV}} (\abs{U_{e1}}, \abs{U_{\mu 1}})}{\mathcal{N}_{\mathrm{events}}^{1-\mathrm{DV}} (\abs{U_{e1}} = 1, \abs{U_{\mu 1}} = 0)} &\simeq \left\lbrace 
    \begin{aligned}
        &\abs{U_{e1}}^2 + \abs{U_{\mu 1}}^2 \\ 
        &\hspace{0.5cm}\text{(for FCC-ee, CEPC)}\,, \\
        &\abs{U_{e 1}}^4\,(\abs{U_{e1}}^2 + \abs{U_{\mu 1}}^2) \\ 
        &\hspace{0.5cm}\text{(for ILC, CLIC)} \,,
    \end{aligned} \right.
\end{align}
and for two DVs,
\begin{align}
    \frac{\mathcal{N}_{\mathrm{events}}^{2-\mathrm{DV}} (\abs{U_{e1}}, \abs{U_{\mu 1}})}{\mathcal{N}_{\mathrm{events}}^{2-\mathrm{DV}} (\abs{U_{e1}} = 1, \abs{U_{\mu 1}} = 0)} &\simeq \left\lbrace 
    \begin{aligned}
        &(\abs{U_{e1}}^2 + \abs{U_{\mu 1}}^2)^2 \\
        &\hspace{0.5cm}\text{(for FCC-ee, CEPC)}\,, \\
        &\abs{U_{e1}}^4\,(\abs{U_{e1}}^2 + \abs{U_{\mu 1}}^2)^2 \\ 
        &\hspace{0.5cm}\text{(for ILC, CLIC)} \,,
    \end{aligned} \right.
\end{align}
and for a muon collider, it would be similar to the case of ILC or CLIC, but with $\abs{U_{e1}}^4 \to \abs{U_{\mu 1}}^4$. 

With all this taken into account, we will consider two different scenarios and compare them with the sensitivity bounds from the main text. We will consider an equal mixing, where $\abs{U_{e1}}^2 = \abs{U_{\mu 1}}^2 = \abs{U_{\tau 1}}^2 = 1/3$ and a the case where the values of the \textit{right-handed} PMNS are equal to the \textit{left-handed} PMNS [this would be true in the so-called \textit{type-II} dominated seesaw in the LRSM, where $M_D$ is negligible in Eq.~\eqref{eq:see_saw_lagrangian}], where $\abs{U_{e1}}^2 \simeq \num{0.68}$ and $\abs{U_{\mu 1}}^2 \simeq \num{0.07}$, for the normal ordering \cite{deSalas:2020pgw, Esteban:2020cvm}.

Figure~\ref{fig:different_mixing_sensitivities} shows how much our sensitivity bounds would change depending on our different mixing scenarios. For FCC-ee the change in sensitivity is not drastic because the main production channel is through a $Z$ boson; for a muon collider, the sensitivity greatly changes because of the diminution in production cross section. For both linear colliders, we would expect a similar decrease in sensitivity as the one from a muon collider.

\section{THE POTENTIAL OF PROMPT DECATS} \label{app:prompt_searches}

As we stated before in the main text, prompt searches also have a very clean signal due to the Majorana nature of HNLs. Indeed, we can have LNV processes in the form of an SS lepton signal and two jets ($\ell\ell jjjj$). The SM has no way of producing any irreducible background to this signal. Possible sources of background arise from heavy hadron decays, as it was said in \cite{Nakajima:2022pkd}. 

A full background analysis is beyond the scope of this paper and should be subject to future study. However, we will perform a very optimistic projection on the possible sensitivity that lepton colliders might have on searches for LNV processes originating from the production of two HNLs. Searches for prompt signals probe a different region of the parameter space than displaced searches, higher values of $m_N$ and smaller values of $M_{W_R}$.

If we only consider LNV-violating processes, we would reduce the signal in half; Majorana HNLs have an equal possibility to decay to both $N \to \ell^- U \bar{D}$ and $N \to \ell^+ \bar{U} D$. The same signal has appeared in searches for the LHC either in the same model \cite{Maiezza:2015lza} or in an extended minimal Type-I seesaw that includes interactions between axions and HNLs \cite{deGiorgi:2022oks}.

We can obtain our bounds in a similar way as we did for the DVs. The expected number of LNV events is
\begin{equation}
    \mathcal{N}_{\mathrm{events}}^{\mathrm{LNV}} \simeq \frac{1}{2}\cdot\mathcal{L}\cdot\sigma(\ell_\alpha^+ \ell_\alpha^- \to NN)\,,
\end{equation}
where the factor $1/2$ comes from the fact that half of the total signal should be LNV. Again, for a backgroundless search, a $2\sigma$ discovery would be equivalent to seeing four events. 

We should also stress that this sensitivity projection comes from using the NWA approximation [which was justified in Eq.~\ref{eq:decay_width_mass_ratio}], and does not take into account the effects that the production of off-shell HNLs might have. The inclusion of off-shell HNLs will also allow us to probe values of $m_N > \sqrt{s}/2$ and will also change the ratio of LNC and LNV processes, as it was shown in \cite{Ruiz:2020cjx}. Of course, we would be expecting the production of off shell HNLs to be suppressed, but it should also be the subject of a future study.

In Fig.~\ref{fig:prompt_sensitivities} we show an estimate of the sensitivity to prompt LNV searches at the FCC-ee with $\sqrt{s}$ at the $Z$ pole and a muon collider with $\sqrt{s} \simeq \SI{10}{TeV}$. 

From the projected sensitivity, we can see that the FCC-ee is best suited for displaced searches. Most of the parameter space that prompt signals would cover would have already been probed by meson-oscillations constraints and from the potential searches done at the HL-LHC for displaced vertices from a single HNL. Prompt searches would allow a muon collider to probe a much bigger region of the parameter space.

With the considerations we have made, our projected sensitivities should not be affected by a non-zero value of $\beta$. But, as we discussed before, we would allow for leptonic decays and decays into an on-shell $W$ boson, whose detection efficiency may change as compared to semi-leptonic signals. All of these considerations should also be taken into account in a future study.

\bibliography{bibliography}
\end{document}